\documentclass{article}
\usepackage{arxiv}
\usepackage[utf8]{inputenc} % allow utf-8 input
\usepackage{url}            % simple URL typesetting
\usepackage{booktabs}       % professional-quality tables
\usepackage{amsfonts}       % blackboard math symbols
\usepackage{nicefrac}       % compact symbols for 1/2, etc.
\usepackage{microtype}      % microtypography
\usepackage{amsmath}
\usepackage{bbm}
\usepackage[ruled,vlined]{algorithm2e}
\usepackage{float}
\usepackage{mathtools}  
\usepackage{amssymb}
\usepackage{tabulary}
\usepackage{amsthm}
\usepackage[titletoc]{appendix}

\usepackage[export]{adjustbox}

\usepackage{graphicx}
\usepackage{chngcntr}
\usepackage{pdfpages}
\usepackage{lineno}
\DeclarePairedDelimiterX\set[1]\lbrace\rbrace{#1}

\usepackage[colorlinks=true,linkcolor=red,filecolor=green,citecolor=blue]{hyperref}
\usepackage{cleveref}
\usepackage{authblk}
\usepackage{tcolorbox}
\usepackage{placeins}
\usepackage{array,xcolor,colortbl}
\usepackage{multirow}

\newcolumntype{G}{>{\centering\columncolor{gray!20!white}}p{0.2\textwidth}}
\newcolumntype{C}{>{\centering\arraybackslash}p{0.2\textwidth}}
\usepackage{verbatim}

\usepackage{titletoc}
    \titlecontents{chapter}
    [0em] %
    {\medskip\bfseries}
    {\thecontentslabel\quad}%numbered chapters
    {}%numberless chapter
    {\hfill\contentspage}

    \titlecontents{section}[1.72em]{\smallskip}%
    {\thecontentslabel.\enspace}%numbered sections
    {}%numberless section
    {\titlerule*[1.2pc]{.}\contentspage}%

\definecolor{colour3}{RGB}{178,55,250} % purple

\newcounter{noteMCctr} 

\newcommand{\bb}{\hspace{0mm}$\bullet$}
\author{Nikolas Michael$^{\dag \S \ast}$, Mihai Cucuringu$^{\dag \ddag \mathparagraph \odot}$ and Sam Howison$^{\ddag}$} 

\affil{
$\dag$Department of Statistics, University of Oxford, 24-29 St Giles', Oxford OX1 3LB, UK\\
        $\ddag$Mathematical Institute, University of Oxford, Andrew Wiles Building, Woodstock Rd, Oxford OX2 6GG \\
        $\mathparagraph$ Oxford-Man Institute of Quantitative Finance, University of Oxford\\
        $\odot$ The Alan Turing Institute, John Dodson House, 96 Euston Rd, London NW1 2DB \\
        $\S$ Boltzmann Research, Interlink Hermes Plaza, 46 Ayiou Athanasiou Avenue, Limassol 3076, Cyprus \\
        $\ast$Corresponding author.
Email: nikolas.michael@kellogg.ox.ac.uk 
        }

\usepackage{natbib}
\usepackage{subfigure} 

\title{A GCN-LSTM Approach for ES-mini and VX Futures Forecasting}

\usepackage{lineno}
\begin{document}

\maketitle

\begin{abstract}
We propose a novel data-driven network framework for forecasting problems related to E-mini S\&P 500 and CBOE Volatility Index futures, in which products with different expirations act as distinct nodes. We provide visual demonstrations of the correlation structures of these products in terms of their returns, realized volatility, and trading volume. The resulting networks offer insights into the contemporaneous movements across the different products, illustrating how inherently connected the movements of the future products belonging to these two classes are. These networks are further utilized by a multi-channel Graph Convolutional Network to enhance the predictive power of a Long Short-Term Memory network, allowing for the propagation of forecasts of highly correlated quantities, combining the temporal with the spatial aspect of the term structure.
\end{abstract}


 {
   \hypersetup{linkcolor=blue}
   \tableofcontents
 }

\section{Introduction and Related Literature} \label{sec:introduction}
\subsection{Futures} 
Futures are highly traded financial instruments used for both risk management and speculation. Compared to directly trading the underlying instruments, many futures allow for 24-hour trading, providing traders with more flexibility. More importantly, as derivative products, they can offer higher leverage, amplifying potential returns and minimizing the cost of hedging. Additionally, they make feasible short-selling a contract, even when the underlying asset is not owned.

In this paper, we focus on the interaction of futures contracts with different expirations. The volatility term structure of VIX futures has been studied by \cite{yingzhi2}, who provide an approximation for the price of VIX futures. A later alternative is \cite{huskaj}, who offer an exogenous model based on specific characteristics. Our goal is to follow a data-driven approach for forecasting the price movement, volatility, and traded volume of these futures by employing data from various products across different expiration periods.

\subsection{SPX and VIX}
We focus on two types of futures: E-mini S\&P 500 futures (ES), with a quarterly cycle, and CBOE Volatility Index futures (VX), with a monthly cycle. Both futures are widely used in describing and tracking the market.

The SPX is a stock market index that measures the performance of 500 of the largest publicly traded companies in the United States, across different industries, making it one of the most widely used gauges of the overall market trend. The VIX represents the market's expectations for volatility over the coming 30 days and reflects investor sentiment on market risk. It is derived from the prices (and, thence, the implied volatilities) of SPX options, as explained in \cite{CBOE_VIX}. High values of VIX indicate increased market uncertainty, while low values suggest a more stable market environment.

The ES futures, traded on the Chicago Mercantile Exchange (CME), are a more accessible alternative to standard S\&P 500 futures, being only one-fifth of their notional size.\footnote{One E-mini futures contract corresponds to 50 times the value of the SPX index, as opposed to 250 times for the standard S\&P 500 futures.} An important question concerns what is the market in which the S\&P 500 price discovery is taking place. While the SPX ETF is generally accepted to lead its derivative products, some studies (e.g., \cite{kurov2004price}) on price discovery have concluded that E-mini futures have a higher informational share than their ETF counterparts during periods with high volatility.

There has also been significant academic interest in the dynamics of the VIX market. \cite{yingzhi} suggested a model for the pricing of VIX futures. \cite{simon2014vix} formulated a simple trading strategy based on whether the basis is in contango (VIX futures are shorted) or in backwardation (VIX futures are longed).\footnote{Contango (Backwardation) occurs when futures prices are higher (lower) than the risk-neutral price for futures.}

This relationship has been shown to be more relevant after major news announcements \cite{chen2017determinants}. Other profitable trading strategies have been suggested, including \cite{TAYLOR20191193}, who used six readily available predictors. \cite{osterrieder} have made use of option pricing data and recurrent neural networks. A causal lead-lag relationship was also established with VIX-tracking Exchange-Traded Products \cite{causality}.

\subsection{Relations between Returns, Volatility and Trading Volume}
Price movements of products tied to the same underlying index are typically highly correlated. While the ES and VX futures track different indices, they can still be informative for each other, as the two indices are naturally intertwined. Specifically, it has been established that returns for VIX and SPX are negatively correlated (see e.g. \cite{Whaley2009}), which is a natural consequence of returns being negatively correlated with volatility. For example, \cite{MOUGOUE1996253} showed this for the German and French stock markets. \cite{BANGSGAARD2024100851} analyzed the lead-lag relationship of futures corresponding to the VIX and SPX indices, showing that under a high-volatility regime, VIX futures lead SPX futures, and the correlation is highly negative.

The relationship between volatility and trading volume has also been widely studied. There is a popular saying that \textit{``it takes volume to move prices"}, which directly explains the positive correlation between the two quantities, with \cite{ahmed} showing how volume is an informative predictor in a volatility model. Theoretical frameworks for this interaction include the mixture of distribution hypothesis by \cite{copeland} and the sequential information arrival hypothesis by \cite{clark}, both modeling how information embedded in trading volume is informative for future changes in price.

There have also been studies on the relationship between returns and volume. For example, \cite{Llorente} showed that the relationship between returns and volume depends on the prevalence of informed trading in the market. In \cite{michael2022option}, we showed how volumes from the option market can be used to forecast future returns for the underlying equities.

\subsection{Graph Neural Networks}
In this paper, we use the methodology of Graph Neural Networks (GNN) to analyze the relationships between different products. GNNs were first introduced by \cite{gori} as a type of neural network capable of dealing with graphical data. They gained rapid popularity after the introduction of Graph Convolutional Networks (GCN) by \cite{kipf2017semisupervised}. These networks and their extensions have been used for many applications, including link prediction \cite{zhang2018link}, node classification \cite{rong2020dropedge}, and outlier detection \cite{Yulei}. Much research has also been conducted on time-series forecasting, including \cite{cao2021spectral}, who introduced the spectral temporal GNN. Traffic forecasting has been a major area of application, with \cite{liyong} and \cite{Djenouri} being major examples of urban and maritime traffic forecasting, respectively.

There have been numerous applications of GNNs in finance. Examples include \cite{matsunaga2019exploring}, who used a knowledge graph to predict stock prices in the Nikkei 225 market, and \cite{li-wei}, who used a combination of LSTM and GCN with news data to classify stocks into upwards and downward trending. More recently, \cite{zhang2023graph} combined GNN with Heterogeneous Auto-regressive Regression models to forecast the volatility and covariance matrix of different instruments. A detailed overview of other GNN applications in finance has been covered in \cite{wang2021review}.

It is important to stress that GNNs encompass quite a diverse class of methods, with many tailored to the specific problem at hand. A multi-layered version of GNN is proposed by \cite{grassia2021mgnn}, who extended GNNs to a multi-layered version by analyzing the intra-layer and inter-layer interactions separately. Furthermore, the time-series nature of a dataset can also be considered using one of the various graph neural network methods tailored for time series forecasting \cite{jin2023survey}. Focusing on research based on stock market forecasting, \cite{research_zhang} also provided a shorter summary of relevant research.

GCNs have also been used by \cite{LI2022108842} to enhance the predictability of chart similarity approaches. The improved graph convolutional network (IGCN) is suggested by \cite{CHEN202167}, which captures both stock market and individual-level features. \cite{xingkun} used an architecture with a GCN and gated recurrent unit (GRU) to show promising performance for stock movement prediction.

\subsection{Main contribution of this paper}
The focus of this paper is the forecasting of three major financial quantities monitored by practitioners—financial returns, volatility, and volume—for two of the major indices (SPX and VIX) that represent the state of the market. The change in the overall market state is captured by the return of SPX, the current and expected volatilities are captured by the volatility of SPX and the value of VIX, while the ``second order volatility" is captured by VIX's volatility. The volumes of instruments corresponding to these indices are deemed to act as a proxy for the market trading activity. A disproportionately high trading activity in VIX could indicate high hedging activity or even a pessimistic view of the market.

While much research has been conducted on the interactions between these quantities, a less explored angle is a data-driven study of the interplay in the futures market with various products of differing expirations. The multitude of products can naturally be represented as a network, where each node of the network represents a different product. This network can act as a visualization of the correlation structure between the different quantities discussed.

More importantly, these networks may be used as input to a graph neural network (GNN). As all of these quantities are highly correlated, a joint forecasting and subsequent propagation via a GNN could prove promising. Such a layer can be used on top of a Long Short-Term Memory (LSTM) network \cite{lstm}, which is a staple in time series forecasting. In later sections, we compare and contrast the base LSTM network as well as other baselines with the GCN-enhanced LSTM network (GCN-LSTM). The version of GCN-LSTM we formulate is suitable for highly correlated graphs. By utilizing features generated by products of differing term structures, we augment and improve the forecasting performance for all three main financial quantities discussed in this paper.

As we are considering different quantities, multiple types of interactions are considered, which can result in different graphs. In particular, a different network is constructed for each pair of variables. These graphs are subsequently used in a multi-channel GCN network to provide parallel propagation attempts at smoothing and improving the base forecasts.

Lastly, as there has been a shift towards loss functions with economic significance, we utilize the Sharpe Ratio in the loss function for financial returns, as well as a quasi-likelihood loss function for volatility.

\paragraph{Paper Structure}
The rest of this paper is structured as follows; In \Cref{sec:model}, we give a high-level overview of the elements of the model that we use. In \Cref{sec:data_construction}, we describe the data set used as well as the graphs that are used by the GCN-LSTM model. In \Cref{sec:setup}, we consider some practical issues, including the evaluation metrics, predictive features, baselines and implementation details. In \Cref{sec:results}, we show the results for the forecasting tasks of returns, volatility and volumes. Finally, in \Cref{sec:conclusion} we conclude and discuss future directions.

\section{Model} \label{sec:model}
\begin{figure}[h]
\centering
   \rotatebox{270}{\includegraphics[trim={0 0 0 0},width=0.52\textwidth]{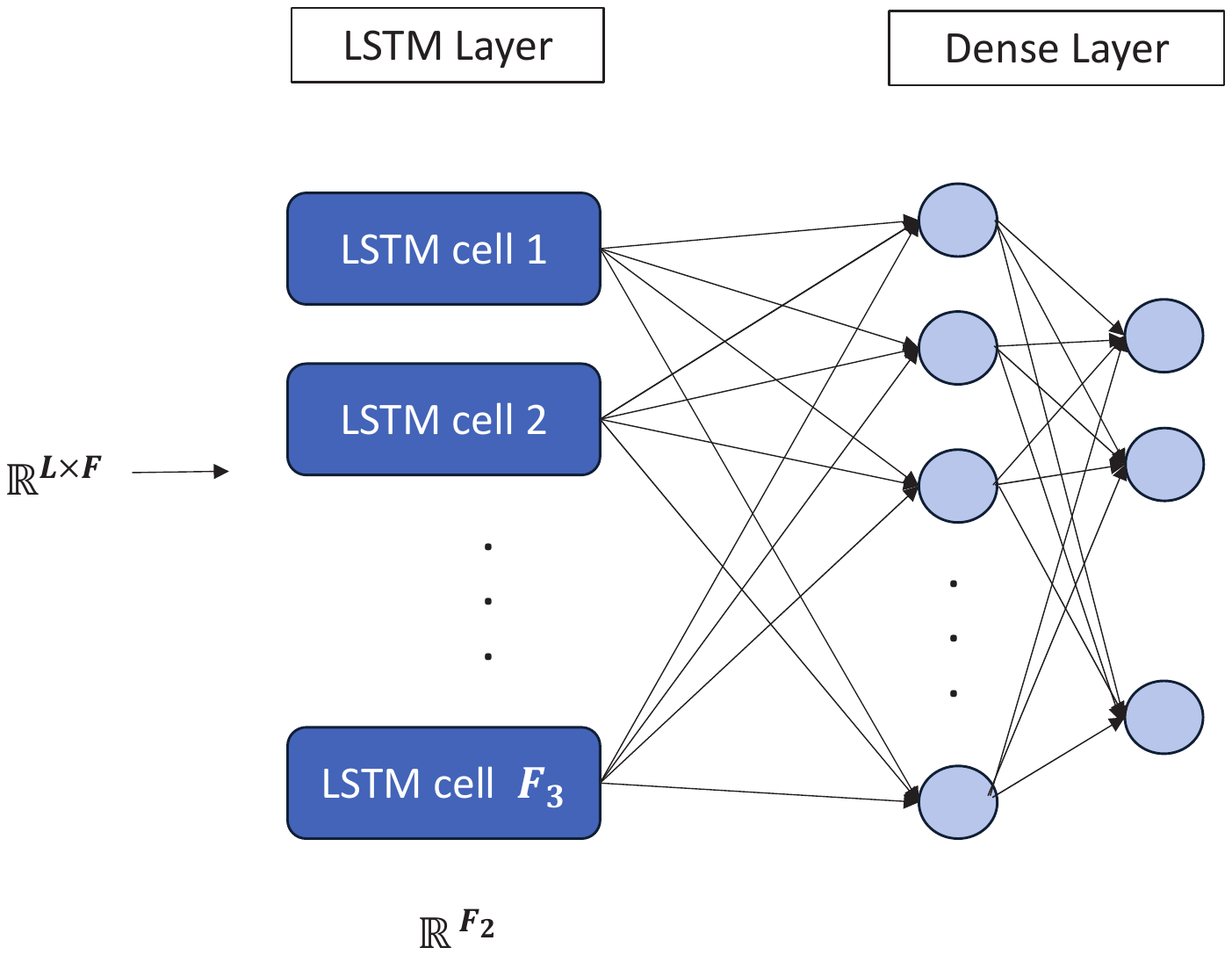}}
   \caption{The module used for each distinct product, comprising an LSTM and a series of dense layers. 
   }
   \label{fig:module_diagram}
\end{figure} 

\begin{figure}[h]
\centering
   \includegraphics[trim={0 70 0 70},width=1\textwidth]{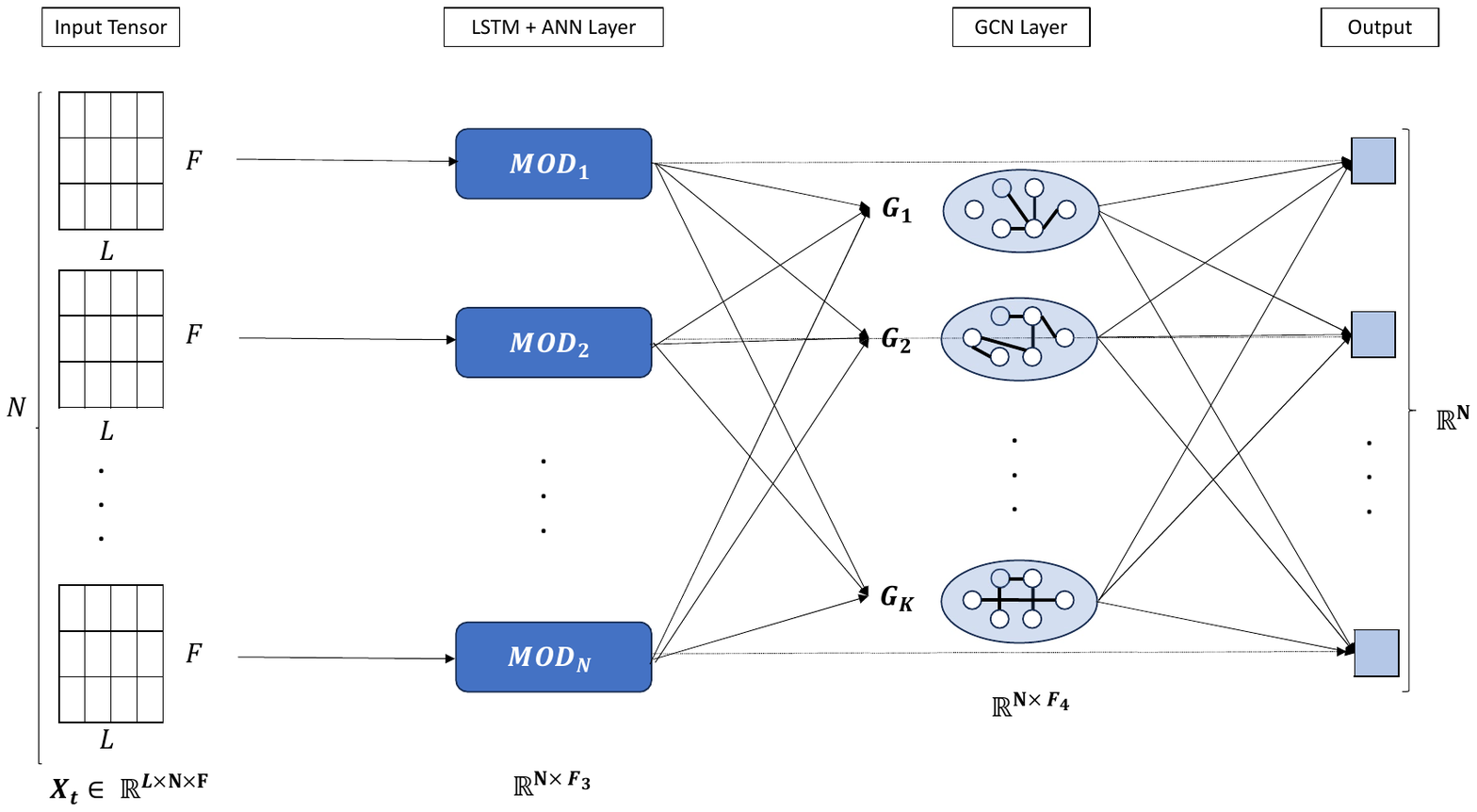}
   \caption{Overview of the GCN-LSTM model used. Each node/instrument corresponds to a separate trainable sub-network,  consisting of LSTM and dense layers as shown in \Cref{fig:module_diagram}. The outputs are then pooled in the GCN layer, which consists of $K$ pre-calculated graphs. A final dense layer is then added to produce the overall output; this also includes skip connections from the individual sub-networks. 
   }
   \label{fig:full_diagram}
\end{figure} 

The model we use for forecasting returns, volatility, and trading volume is an iteration of the GCN-LSTM model. This approach involves combining an LSTM network layer, which is widely used for datasets with a temporal structure, such as the financial data in our study. For each individual product, a distinct module operates in parallel, as illustrated in \Cref{fig:module_diagram}. 

Although this increases the computational resources required and reduces the number of available trainable data points, it enables the model to assign differing weights to each product, thereby enhancing its expressive power. As shown in \Cref{fig:full_diagram}, the architecture incorporates a GCN layer that integrates multiple graphs, allowing the individual forecasts for different products to be used in conjunction. Finally, the GCN layer is followed by a dense layer, which includes skip connections from the last dense layer before the GCN layer.
In the remainder of this section, 
we provide an overview of the different layers employed. 

\subsection{Graph Convolutional Networks}
Graph Neural Networks (GNNs) represent a class of deep learning methods designed for network data, based on the concept of neural message passing. Following the notation of \cite{hamilton}, we consider a set of nodes $V = \{ v_1,\dots,v_N \}$. We denote by $\mathcal{N}(v)$ the set of nodes that are connected with node $v$. 

At each layer $k$, the message of node $v$, represented as a vector of length $F_k$, is denoted by $h_v^{k}$ and is calculated as
\begin{equation} \label{eq:update}
    h_v^{k} = \textrm{UPDATE} \left( h_v^{(k-1)}, m_{\mathcal{N}(v)} \right),
\end{equation}  
where \begin{equation} \label{eq:aggregate}
    m_{\mathcal{N}(v)} = \textrm{AGGREGATE} \left( \{ h_u^{(k-1)}, \forall u \in \mathcal{N}(v) \}\right).
\end{equation}
We choose the AGGREGATE operation to be the mean function, i.e the mean message coming from a node's neighbors.
For single-layered networks, this structure allows each node to be promptly updated using information from its neighbors. Like other neural network architectures, multiple such layers can be stacked on top of each other for higher-order propagation. The result is that deeper networks can incorporate information from nodes that are at a higher geodesic distance from the current node.

One of the most widely used iterations of GNN is the Graph Convolutional Network (GCN), initially proposed by \cite{kipf2017semisupervised}. In its simplest form, a single layer of GCN can be expressed as
\begin{equation} \label{eq:gcn_simple}
    h_v = \sigma \left( W \sum_{z \in \mathcal{N}(v) \cup \{ v \}} \frac{h^0_z}{\sqrt{ |\mathcal{N}(v)| \, |\mathcal{N}(u)|}} \right), 
\end{equation} 
where $h_z^0$ is the input vector for node $z$, $W$ is the $F_1 \times F$ weight matrix, and $\sigma$ is an activation function (we use the tanh function). Given that we are dealing with a network of just 14 nodes, a single-layer architecture using an UPDATE function like \Cref{eq:gcn_simple} is sufficient for effectively propagating forecasts across different products, particularly one that allows for weighted and signed graphs.

In our case, we are dealing with multiple distinct graphs, which necessitates training different weight matrices along with the rest of the network. Denote by $A^{(i)}$ the adjacency matrix of the $i^{\textrm{th}}$ graph, by $D^{(i)}$ the $N\times N$ diagonal matrix whose $i^{\textrm{th}}$ entry is equal to $|\mathcal{N}(v_i)|$, and by $I_N$ the $N\times N$ identity matrix. Also, denote by $H^{(0)}$ the $N\times F$ matrix whose $i^{\textrm{th}}$ column is equal to $h^0_z$. The output of this layer is thus an order-3 tensor $\left( H^{(1)},\dots,H^{(K)} \right)$ computed as
\begin{equation}  \label{eq:H}
    H^{(i)} = \textrm{ReLU}  \left( (D^{(i)})^{-1/2} \left( 
 A^{(i)} - I_N \right) (D^{(i)})^{-1/2}  H^{(0)}W^{(i)} \right), \qquad  i=1,\dots,K, 
\end{equation}  
where $W^{(i)}$ are the $F_1 \times F$ matrices to be optimized. 

\subsection{Long-Short Term Memory Networks (LSTM)}
Recurrent Neural Networks (RNNs) form a class of neural networks specifically designed for datasets with temporal structures. They achieve this by retaining information from the processing of previous time steps, which is used in conjunction with the input of the current time steps. The architecture and parameters are shared across time steps.

A common issue with simple RNNs is that gradients have a tendency to either vanish or explode during training. Long Short-Term Memory (LSTM) networks \cite{lstm} are an RNN variant specifically designed to mitigate these issues by introducing a cell state. This cell state contains a summary of the previous cells and is regulated by several gates that improve gradient flow.

Denote by $x_t$, $c_t$, and $h_t$ the input, cell state, and output of the module corresponding to time point $t$, respectively. The formula of a basic LSTM unit takes the form:
\begin{equation} \label{eq:lstm}
\begin{alignedat}{2}
& \textrm{Forget Gate: } & f_t &= \sigma(W_f x_t + U_f h_{t-1} + b_f), \\
& \textrm{Output Gate: } & o_t &= \sigma(W_o x_t + U_o h_{t-1} + b_o), \\
& \textrm{Input Gate: } & i_t &= \sigma(W_i x_t + U_i h_{t-1} + b_i), \\
& \textrm{Candidate Cell State: } & \hat{c}_t &= \textrm{tanh}(W_c x_t + U_c h_{t-1} + b_c), \\
& \textrm{Updated Cell State: } & c_t &= f_t \odot c_{t-1} + i_t \odot \hat{c}_t, \\ 
& \textrm{Hidden State: } & h_t &= c_t \odot \textrm{tanh}(c_t).
\end{alignedat}
\end{equation}

where $W_f,U_f,b_f,W_i,U_i,b_i,W_o,U_o,b_o,W_c,U_c$ and $b_c$ are all trainable parameters, and $\odot$ is the Hadamard product. 

\section{Data and Graph Construction} \label{sec:data_construction}
\subsection{Data} \label{sec:data}

We use Level-1 data provided at tick-level for E-mini $S\&P$ 500 and VX futures. At each point, we consider the eight VX futures and four ES futures closest to expiration, respectively. We also include the SPX and VIX indices in our analysis. We denote by VX\_1 (ES\_1) the current VX (ES) futures contract that is closest to expiration, by VX\_2 (ES\_2) the VX (ES) futures contract that is second-closest to expiration, and so on. The corresponding indices are denoted as VIX and SPX.

The time period considered spans 10 years, specifically from January 2014 to December 2023. We remove any entries that correspond to zero liquidity on either the bid or ask side. Our focus is on the hourly level, forecasting hourly returns, realized volatility, and a proxy for trading volume. Specifically, as we do not have access to trading volume data, we approximate the number of trades by counting the number of updates to the top of the book data caused by trades (excluding cancellations and new orders).\footnote{While different from volume, the proxy we use is closely related to volume. This is because a very large proportion of such trades appear to correspond to trading volumes of one contract.}

\begin{figure}[h]
   \includegraphics[width=1\textwidth]{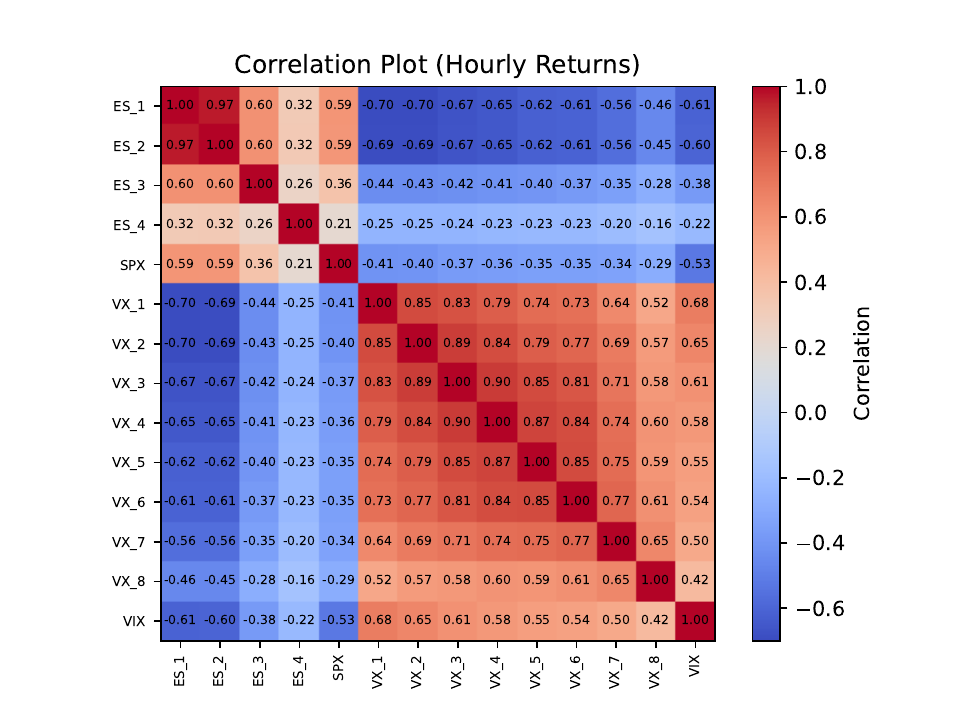}
   \caption{(Spearman) Correlation plot based on the hourly returns of ES and VX futures. 
   }
   \label{fig:corelation_plot}
\end{figure}

A plot of Spearman correlations of hourly returns is shown in \Cref{fig:corelation_plot}. ES futures clearly exhibit high positive correlations with each other, as well as with the SPX index, a pattern also observed for VX futures products. Naturally, products with similar expirations tend to have higher correlations. We also observe a negative correlation between ES and VX futures, with near-month ES (VX) contracts generally having higher correlations with VX (ES) futures compared to contracts with longer expiration periods.

\subsection{Graph Construction}
\label{sec:construction}
Similar to \cite{Shi2024}, we use multiple graphs as input for the GCN layer. The constructed graphs are signed (i.e in the sense that the edges may take either positive or negative weights) and based on different ordered pairs of the three main quantities: returns, volatility, and trading volume. For each ordered pair of variables, we build four different adjacency matrices using Spearman Correlation; Contemporaneous Weighted, Contemporaneous Unweighted, Lagged Weighted and Lagged Unweighted. 

We calculate $C_{x,y}$ as the Spearman correlation between vectors $x$ and $y$. When constructing one of the \textbf{Contemporaneous} graphs, we use vectors $x$ and $y$ that are aligned (i.e elements at the same index correspond to the same time period). The lagged correlation, with a lag of one, is calculated when constructing one of the \textbf{Lagged} graphs, i.e we use $\left( x_1, \dots, x_{T-1} \right)$ and $\left( y_2, \dots, y_{T} \right)$ instead of $x$ and $y$.

Since we are dealing with a small number of nodes whose attributes are all highly correlated, we build the adjacency matrices to avoid working with a complete graph. A natural solution is to use a weighted adjacency matrix based on the correlation matrix. Except for the diagonal entries, which are set to 1, all positive and negative entries are normalized separately column-wise, as shown below.
\begin{equation} \label{eq:weighted}
A^{(\textrm{Weighted})}_{v_i,v_j}  = \mathbbm{1}_{\left( i  = j \right)} +  \mathbbm{1}_{\left( i  \ne j, C_{v_i,v_j} > 0 \right)} \frac{C_{v_i,v_j}}{\sum\limits_{j \ne i} C_{v_i,v_j} \mathbbm{1}_{\left( C_{v_i,v_j} > 0 \right)}} +  \mathbbm{1}_{\left( i  \ne j, C_{v_i,v_j} < 0 \right)} \frac{C_{v_i,v_j}}{\sum\limits_{j \ne i} \left( -C_{v_i,v_j} \right) \mathbbm{1}_{\left( C_{v_i,v_j} < 0 \right)}}.
\end{equation}
We also consider an alternative method to produce an unweighted graph. Given the high correlation among most instruments, the resulting graph would nearly be a complete graph. To address this, we limit the number of edges by including only a fixed number of connections, specifically the $K$-nearest neighbors to each node.
\begin{equation}
A^{(\textrm{Unweighted})}_{v_i,v_j}  = \mathbbm{1}_{\left( i  = j \right)} +  \mathbbm{1}_{\left( i  \ne j, i \in \textrm{NN}_{K}\left( j \right)\right)}  \textrm{sign} \left( C_{v_i,v_j} \right), 
\end{equation}
where 
\begin{equation}
\textrm{NN}_{K}(k) = \biggl\{ j: \Big|  \bigl\{ i \ne k : \left| C_{i,k} \right| \ge \left| C_{j,k} \right|  \bigr\} \Big| \le K \biggr\}. 
\end{equation}

\subsection{Discussion}
\label{sec:analysis_graphs}
\begin{figure}[htbp]
    \centering
        \subfigure[Returns-Returns]{\includegraphics[trim=2.2cm 1.7cm 2.2cm 1.8cm, clip, ,width=0.45\textwidth]{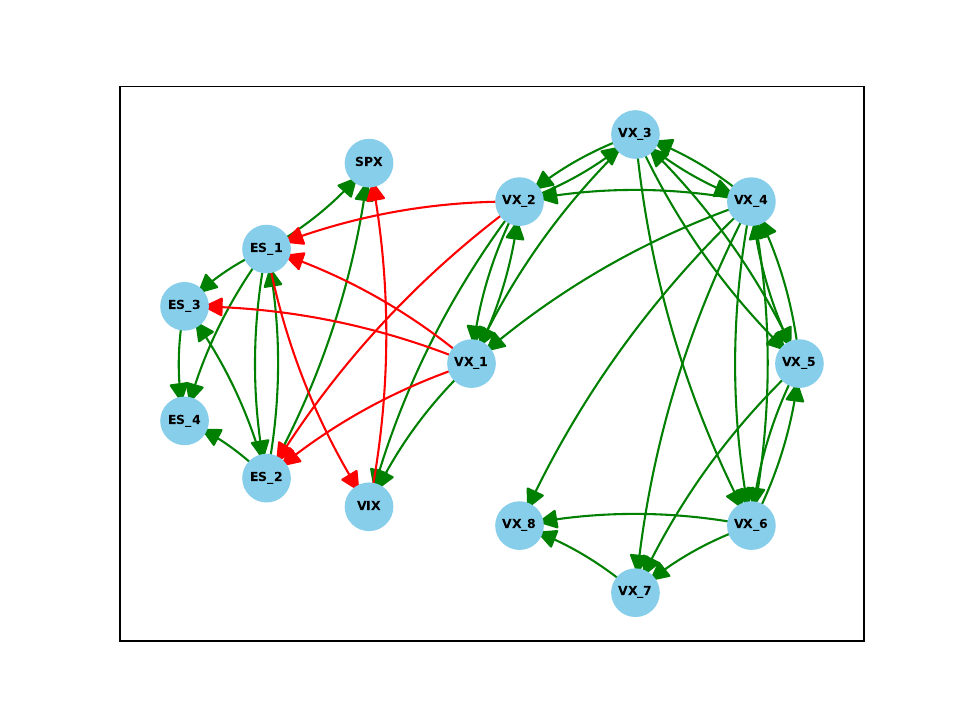}}
    \hfill
    \subfigure[Returns-Volatility]{\includegraphics[trim=2.2cm 1.7cm 2.2cm 1.8cm, clip, ,width=0.45\textwidth]{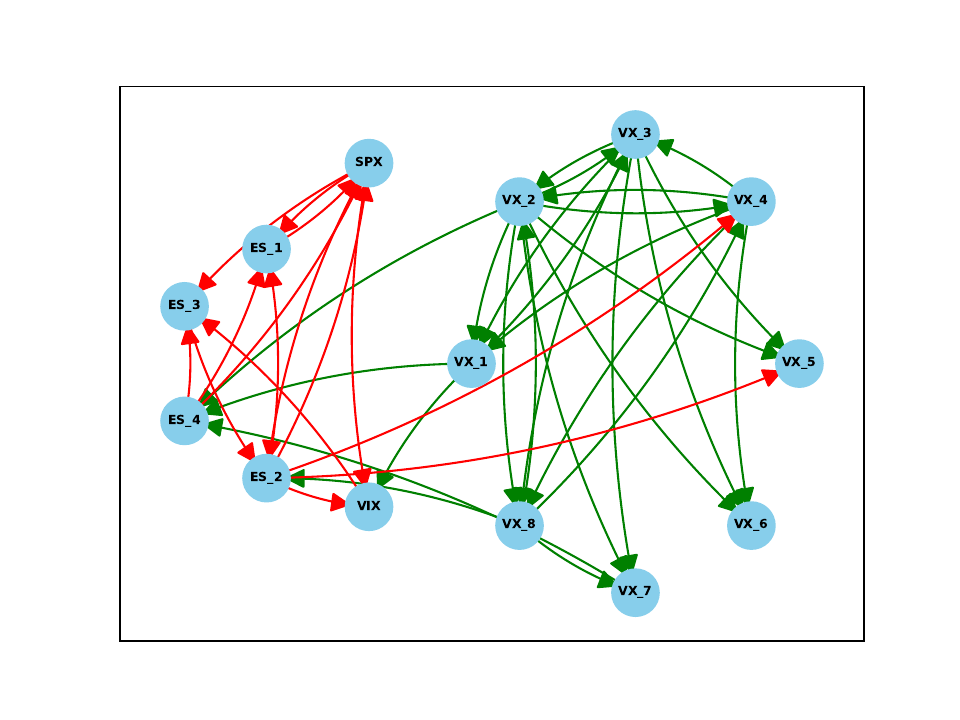}}
    \vspace{0.5cm}
    \subfigure[Returns-Volume]{\includegraphics[trim=2.2cm 1.7cm 2.2cm 1.8cm, clip, ,width=0.45\textwidth]{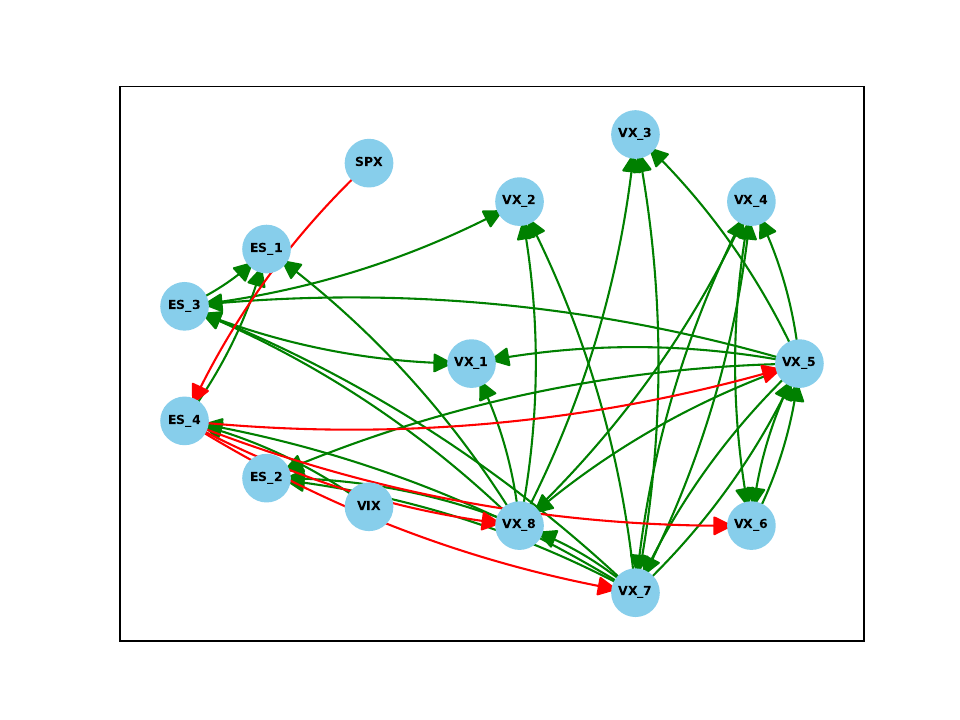}}
    \hfill
    \subfigure[Volatility-Volatility]{\includegraphics[trim=2.2cm 1.7cm 2.2cm 1.8cm, clip, ,width=0.45\textwidth]{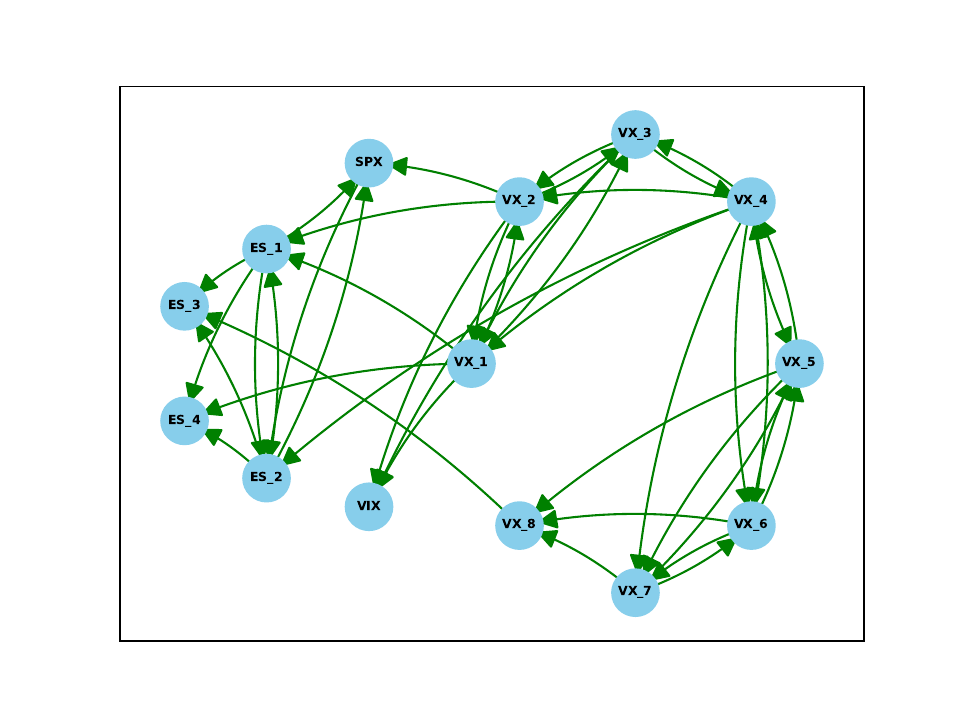}}
    \vspace{0.5cm} % Adjust vertical spacing between rows
    \subfigure[Volatility-Volume]{\includegraphics[trim=2.2cm 1.7cm 2.2cm 1.8cm, clip, ,width=0.45\textwidth]{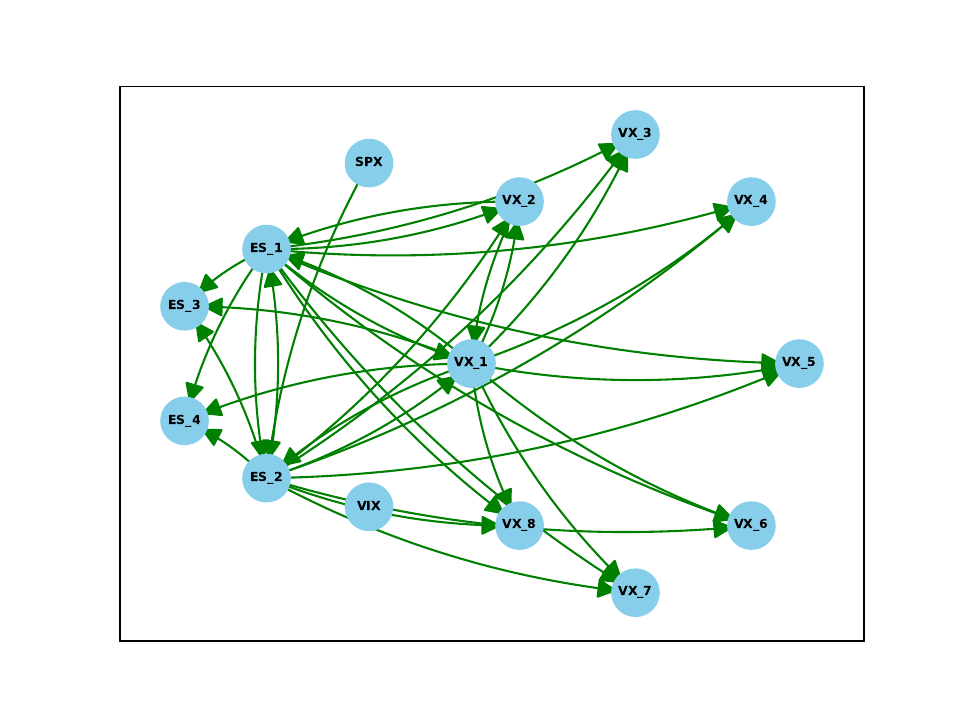}}
    \hfill
    \subfigure[Volume-Volume]{\includegraphics[trim=2.2cm 1.7cm 2.2cm 1.8cm, clip, ,width=0.45\textwidth]{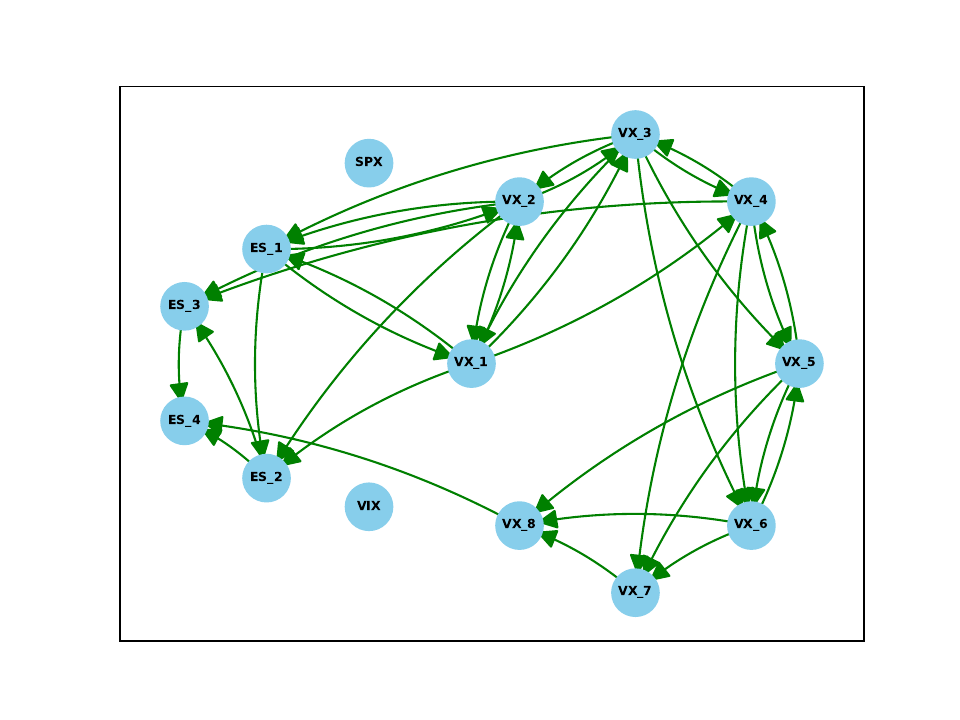}}
    \caption{Visualization of networks. The networks shown are signed, unweighted, and directed. Green arrows correspond to positive edges, while red corresponds to negative edges.}
    \label{fig:graph_list}
\end{figure}

\begin{figure}
    \centering
      \includegraphics[width=\textwidth]{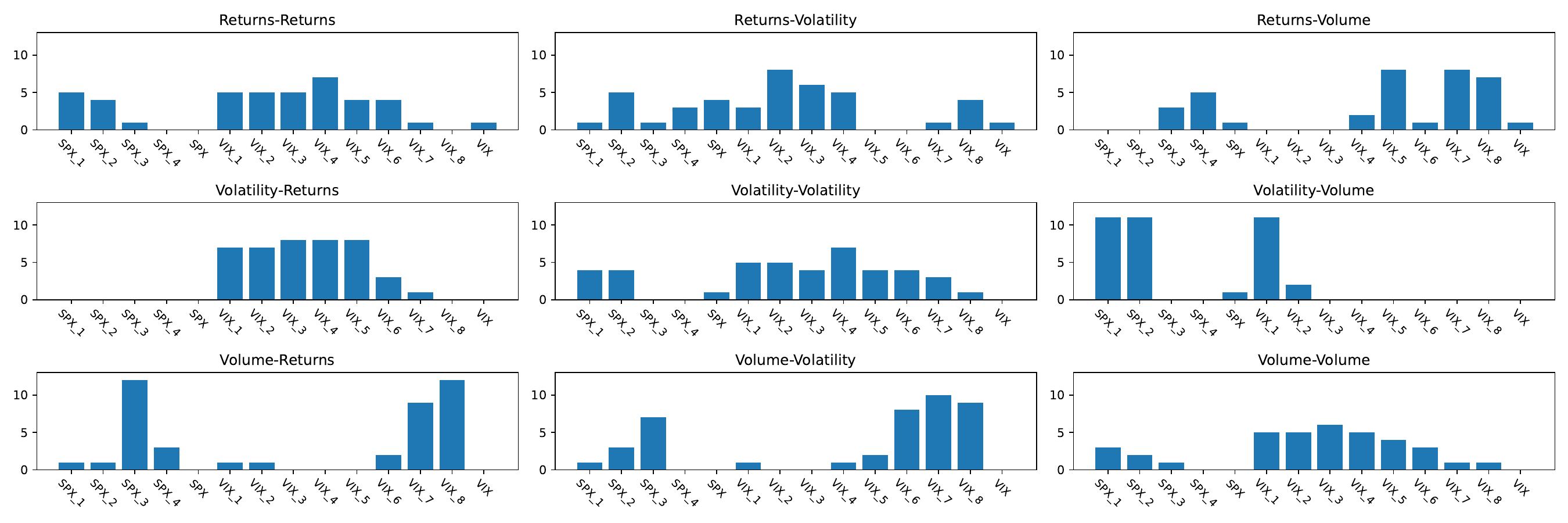} \caption{Barplot for the out degrees of the contemporaneous unweighted graphs.} \label{fig:grid_degrees} 
\end{figure}

In \Cref{fig:graph_list}, we show a selection of the Contemporaneous Unweighted graphs (as discussed in \Cref{sec:results}, Contemporaneous graphs tend to outperform Lagged graphs). For brevity, we only include one graph for each pair, as the graphs corresponding to the two ordered pairs of a pair are based on the same correlation matrix (but transposed). It is also important to note that since indices do not have corresponding trading volumes, there are no outgoing (incoming) edges for graphs where the first (second) pair variable is volume. In \Cref{fig:grid_degrees}, we present a bar plot of the out-degrees for these graphs, while in \Cref{tab:network_details}, we display some important summary statistics regarding these graphs.

The structure of most of these graphs is generally determined by the two product clusters (SPX and VIX). A few graphs, such as the Return-Return and Volatility-Volatility graphs, are highly clustered with only a few connections between the two clusters, most of which go from the VIX cluster to the SPX cluster and not vice versa. This is partly due to the fact that multiple VIX futures products have high liquidity, in contrast to SPX futures, where most liquidity is concentrated in ES\_1.

For the graphs where return is one of the pair variables, there are multiple negative edges across nodes, while no such edges exist for the other graphs. This is largely due to the negative correlation between the market's return and volatility, which is evident from the large number of negative edges between the VIX and SPX clusters in the Return-Return graph. This phenomenon is also observed in the Return-Volatility graphs, where all edges between SPX nodes are negative. Interestingly, SPX nodes are mostly connected to other SPX nodes, despite VIX returns and SPX volatility being highly correlated. In the Volatility-Return graph, there are no edges originating from SPX nodes, with edges between VIX nodes being positive and those from VIX to SPX being negative. Lastly, in the Volume-Return graph, the edges from VIX to SPX are also negative, while the edges from SPX to VIX are positive. Interestingly, ES\_3 has positive edges toward all VIX-related nodes.

It is noteworthy that in certain types of edges within the SPX cluster, and some cross-cluster edges, higher connectivity is shown by more liquid products (ES\_1, ES\_2, VX\_1, and VX\_2), while in other cases, products with very limited liquidity are more connected (ES\_3, ES\_4, VX\_7, and VX\_8). The Return-Return, Volatility-Volatility, and Volume-Volume graphs fall into the former category, as they are all based on high-correlation matrices. Another such network is Volatility-Volume, possibly due to the fact that the prices of these high-liquidity products are less prone to outliers, making large fluctuations possible only when the overall market trading volume is high.

More interesting are the cases where low-liquidity products play a bigger role. As can be seen in \Cref{fig:grid_degrees}, for Returns-Volume, Volume-Returns, and Volume-Volatility, the low-liquidity nodes such as ES\_3, VX\_7, and VX\_8 have very high out-degrees. In particular, for Returns-Volume, VX\_8 seems to have positive edges towards most SPX products, while ES\_4 appears to have negative edges towards many VIX nodes.\footnote{It should be noted that this pair of variables has a graph based on very low correlations (see \Cref{tab:network_details}), which could make the resulting graph less reliable.} For Volume-Volatility, most connections from VIX to SPX occur with VX\_7 and VX\_8.

Note that a different pattern holds for edges connecting two VIX products, where products with expirations that are close tend to be connected. This results in a more uniform distribution of out-degrees for this part of the network, with VX\_3 and VX\_4 being the nodes with the highest (positive) out-degrees towards the VIX cluster in graphs like Return-Return, Volatility-Volatility, Volume-Volume, and Volatility-Volume.

No particular patterns are observed for the Volume-Volume graph, which partly stems from the fact that this graph is derived from a matrix where most correlations are quite high. A few other interesting observations include ES\_3 returns having a high negative correlation with some VIX products' volatility; and that for Volume-Volatility, most edges seem to originate from VIX nodes, with high maturities having a large out-degree (not shown in \Cref{fig:graph_list}).

Overall, while SPX and VIX tend to form the two primary clusters in most of these graphs, there is still a significant number of edges between SPX and VIX nodes. Additionally, while higher-liquidity products such as ES\_1 and VX\_1 have high connectivity, medium and low-liquidity products are also highly connected in certain graphs, like the Return-Volume and Volume-Volatility graphs. While index nodes still have some connections, they are generally among the least connected (as can also be seen in \Cref{fig:grid_degrees}). Altogether, these observations suggest that a GCN layer would be beneficial, as various interactions exist between nodes that cannot be attributed to a single rule, but rather hint at a complex set of interactions underpinning the dynamics of the market.

\begin{table}[]
\resizebox{\textwidth}{!}{
\begin{tabular}{c|cc|cc|cccc|c}
                        & \multicolumn{2}{c|}{Number of Edges} & \multicolumn{2}{c|}{Highest Out Degree} & \multicolumn{4}{c|}{Number of Edges between clusters} & Quantiles                    \\
                        & Positive          & Negative         & Positive            & Negative          & SPX to SPX  & SPX to VIX  & VIX to SPX  & VIX to VIX  & (0,25,50,75,100)             \\ \hline
Return - Return         & 35                & 7                & 7 (VX\_4)          & 3 (VX\_1)        & 9           & 1           & 6           & 26          & (-0.69,-0.40,0.22,0.66,0.96) \\
Return - Volatility     & 27                & 15               & 8 (VX\_2)          & 5 (ES\_2)        & 10          & 4           & 5           & 23          & (-0.10,-0.03,0.01,0.04,0.13) \\
Return - Volume         & 31                & 5                & 8 (VX\_5)          & 4 (ES\_4)        & 3           & 6           & 9           & 18          & (-0.04,0.00,0.00,0.01,0.03)  \\
Volatility - Return     & 27                & 15               & 6 (VX\_3)          & 4 (VX\_4)        & 0           & 0           & 15          & 27          & (-0.10,-0.03,0.01,0.04,0.13) \\
Volatility - Volatility & 42                & 0                & 7 (VX\_4)          & NA                & 9           & 0           & 6           & 27          & (0.01,0.11,0.30,0.45,0.88)   \\
Volatility - Volume     & 36                & 0                & 11 (ES\_1)         & NA                & 7           & 16          & 5           & 8           & (0.00,0.05,0.09,0.17,0.31)   \\
Volume - Return         & 30                & 12               & 9 (ES\_3)          & 4 (VX\_8)        & 5           & 12          & 10          & 15          & (-0.04,0.00,0.00,0.01,0.03)  \\
Volume - Volatility     & 42                & 0                & 10 (VX\_7)         & NA                & 4           & 7           & 11          & 20          & (0.00,0.05,0.09,0.17,0.31)   \\
Volume - Volume         & 36                & 0                & 6 (VX\_3)          & NA                & 4           & 2           & 8           & 22          & (0.00,0.00,0.61,0.85,0.98)  
\end{tabular}} \caption{Summary of the different networks for the nine different graphs for the unweighted contemporaneous graphs.} \label{tab:network_details} 
\end{table}

\section{Forecasting Setup} \label{sec:setup}

\subsection{Financial Terms}
Denote by $P^{\left(\textrm{B} \right)}_t 
\left(\mbox{resp. }  P^{\left( \textrm{A} \right)}_t \right)$ the bid (resp. ask) price. The mid-price $P^{\left(\textrm{M} \right)}_t$ is calculated as 
\begin{equation} \label{eq:midprice}
P^{\left(\textrm{M}\right)}_t = \frac{P^{\left(\textrm{A}\right)}_t + P^{\left(\textrm{B}\right)}_t}{2}.
\end{equation} 
The $k$-minute return observed at time $t$ is then defined as
\begin{equation} \label{eq:R_minute}
R_t^{k\text{-minute}} = \frac{P^{\left(\textrm{M}\right)}_t -  P^{\left(\textrm{M}\right)}_{t - k}}{P^{\left(\textrm{M}\right)}_{t - k}} \approx \log \frac{P^{\left(\textrm{M}\right)}_t}{P^{\left(\textrm{M}\right)}_{t - k}}.
\end{equation} 
Suppose that $\hat{R}_t^{k-\textrm{minute}}$ is a forecast of $R_t^{k\text{-minute}}$, acting as a trading signal. The forecasts are used as signals for a trading strategy, as a way of evaluating its performance. Specifically, if the sign of $\hat{R}_t^{k\text{-minute}}$ is positive, a theoretical buy transaction is conducted, while a theoretical sell transaction is conducted when the sign is negative. 

Consider $D_d$ as the set containing the time indices corresponding to day $d$. The daily Profit and Loss (P\&L) for day $d$ is given by 
\begin{equation} \label{eq:PNL}
P \& L_d = \sum_{t \in D_d} R_t^{k\text{-minute}} \textrm{sign} \left( \hat{R}_t^{k\text{-minute}}\right).
\end{equation}   
The annualized Sharpe Ratio is calculated as 
\begin{equation} \label{eq:SR}
\textrm{SR} = \frac{\textrm{mean}(\{R_d\}_{d=1}^D) \times \sqrt{252}}{\textrm{sd}(\{R_d\}_{d=1}^D)},
\end{equation}   

The Profit Per Dollar traded is calculated as
\begin{equation} \label{eq:PPD}
    \textrm{PPD} = \textrm{mean}(\{ P \& L_d \}_{d=1}^D). 
\end{equation} 
We also define the k-minute realized variance calculated as 
\begin{equation} \label{eq:RV_hourly}
    {RV}^{k\text{-minute}}_t = \sum_{j=0}^{K-1} \left(R_{t-j}^{1\text{-minute}} \right)^2.
\end{equation} 
As realized volatility is highly skewed, we instead forecast $\text{log} \left({RV}^{k\text{-minute}}_t  \right)$.
The $k$-minute volume at time $t$ is denoted as $V_t^{k\text{-minute}}$, and is calculated as the number of updates in the orderbook caused by trades, that occurred up to $k$ minutes prior to $t$. 
Note that trading volume is not a stationary time-series, as it generally increases over time. Therefore, we forecast the percentage change in volume defined as 
\begin{equation} \label{eq:delta_vol}
    \delta_{t}^{k\text{-minute}} = \log \frac{V_t^{k\text{-minute}}}{V_{t-K}^{k\text{-minute}}}.
\end{equation}
\subsection{Loss functions and evaluation metrics}
Denote by $\{ y_t \}_{t=1}^T$ the labels of the true values to be forecasted (in our case this is either the returns, realized volatility or trading volume), and by $\{ \hat{y}_t \}_{t=1}^T$ the forecasts given by the model. 
When forecasting trading volume, we use the Mean Absolute Error loss function
\begin{equation} \label{eq:mae}
    \textrm{MAE} = \frac{1}{T} \sum_{t=1}^T \left| y_t - \hat{y}_t \right|.
\end{equation}  
We also use the Mean Squared Error which is equal to 
\begin{equation} \label{eq:MSE}
    L_{\textrm{MSE}} = \frac{1}{T} \sum_{t=1}^T (y_t - \hat{y}_t)^2.
\end{equation} 
In contrast to trading volume, financial returns are intrinsically much harder to forecast. Furthermore, typical loss functions, such as the MSE, penalize predictions that have a large deviation. However, when forecasting returns, predictions that are correct in sign but wrong in magnitude are much less detrimental than those that are wrong in sign. Therefore, using a variable with economic significance would be a sensible approach. While the Sharpe Ratio would be a natural choice, it is not differentiable with respect to $\hat{y}_t$ due to the usage of the $\textrm{sign}$ function. To circumvent this, let $\epsilon > 0$ be a very small constant. We consider

\begin{equation} \label{eq:SR_loss}
    L_{\textrm{SR}} = - \frac{\textrm{mean}(\{ \tilde{R}_t \}_{t=1}^T)}{\left( \textrm{sd}(\{ \tilde{R}_t \}_{t=1}^T) + \epsilon \right)}, 
\end{equation}
where
\begin{equation} \label{eq:R}
    \tilde{R}_t = y_t \textrm{tanh} \left( \frac{1}{\epsilon} \hat{y}_t \right). 
\end{equation} 
\Cref{eq:SR_loss} 
is a slight simplification of the negative annualized Sharpe Ratio, but it is also differentiable, as it does not make use of the $\textrm{sign}$ function. The loss function we use for returns is a combination of these two loss functions
\begin{equation} \label{eq:mixedloss}
    L_{\textrm{mixed}} = L_{\textrm{MSE}} + \alpha L_{\textrm{SR}},
\end{equation} 
where $\alpha$ is another hyperparameter to be chosen. Similar loss functions have been used before, for example  by \cite{Zhang8}, \cite{michael2022option} and \cite{fingan}.  

We also use a custom Quasi-Likelihood (QLIKE) loss function for volatility, defined as
\begin{equation} \label{eq:qlike}
    L_\textrm{QLIKE} = \frac{1}{T} \sum_{t=1}^T \frac{\exp{y_t}}{\exp{\hat{y}_t}} - (y_t- \hat{y}_t) - 1,
\end{equation}  
where the $-1$ term can be omitted for training. QLIKE works independently of the volatility's scale. QLIKE is also used for evaluation of the results, alongside the Heteroskedasticity Adjusted Mean Squared Error (HMSE), defined as 
\begin{equation} \label{eq:hmse}
    \textrm{HMSE} = \frac{1}{T} \sum_{t=1}^T \left( 1 - \frac{\exp{\hat{y}_t}}{\exp{y_t}} \right)^2.
\end{equation}  
Note that both QLIKE and HMSE have both been assessed to be robust measures for volatility forecasting (e.g by \cite{PATTON2011246}). 

\subsection{Predictive Features}
In the following paragraphs, we list the features used in our predictive models. Most of these are variations of predictors that are widely used by practitioners. 
Denote by $P^{(B)}_t$ (resp., $P^{(A)}_t$) the top-of-book bid (resp., ask) price at time $t$. Also, denote by $q^{(B)}_t$ ($q^{(A)}_t$) the top-of-book bid (ask) volume, respectively.

We consider the following features 
\begin{itemize}
    \item The Order-Flow Imbalance (OFI) formulated by \cite{Cont_2013} is used for 5, 10, 30, 60, 90, 120, 180, 270 and 360 minutes. Denote by $E(t,k)$ the set containing the events that occur between time point $t$ and up to $k$ minutes before it. Then, the contribution of each event in $E(t,k)$ calculated as
    \begin{equation} \label{eq:e_for_OFI}
        e_n = \mathbbm{1} \left( P^{(B)}_n \ge P^{(B)}_{n-1} \right) q^{(B)}_n - 
      \mathbbm{1} \left(P^{(B)}_n \le P^{(B)}_{n-1} \right) q^{(B)}_{n-1} - \mathbbm{1} \left(P^{(A)}_n \le P^{(A)}_{n-1} \right) q^{(A)}_{n} + \mathbbm{1} \left(P^{(A)}_n \ge P^{(A)}_{n-1} \right) q^{(A)}_{n-1}.
    \end{equation}  
    The OFI is then computed as
    \begin{equation} \label{eq:OFI} 
            \textrm{OFI}_t^{(k)} = \sum_{n \in E(t,k)} e_n. 
    \end{equation}  
    \item $k\text{-minute}$ returns and realized volatility of the preceding 5, 10, 30, 60, 90, 180 and 390 minutes, computed using \Cref{eq:R_minute,eq:RV_hourly}.
\item $k\text{-minute}$ positive and negative realized semivolatility of the preceding 5, 10, 30, 60, 90, 180 and 391 minutes, computed as 
\begin{equation}  \label{eq:rv_pos}
    \textrm{RV-pos}_t^{k\text{-minute}} = \sum_{j = 0}^{K-1} \left( R_{t-j}^{1-minute} \right)^2 \mathbbm{1}\left( R_{t-j}^{1-minute} \ge  0 \right). 
\end{equation}   
\begin{equation} \label{eq:rv_neg}
    \textrm{RV-neg}_t^{k\text{-minute}} = \sum_{j = 0}^{K-1} \left( R_{t-j}^{1\text{-minute}} \right)^2 \mathbbm{1}\left( R_{t-j}^{1\text{-minute}} \le  0 \right).
\end{equation}
    \item Exponentially weighted return and realized volatility with weight $w \in \{0.75,  0.9, 0.975, 0.99, 0.999, 1\}$ computed as 
    \begin{equation}  \label{eq:r_weighted}
        R_t^{(w)\text{-Weighted}} = (1-w)  \sum_{j=0}^{1440} w^j R^{1\text{-minute}}_{t-j}, 
    \end{equation} 
    \begin{equation} \label{eq:rv_weighted}
        \textrm{RV}_t^{(w)\textrm{-Weighted}} = (1-w)  \sum_{j=0}^{1440} w^j \left( \textrm{RV}^{1\text{-minute}}_{t-j} \right)^2.
    \end{equation}  
    \item Delta-volume, based on the number of trades occurred in the last 5, 10, 30, 60, 90, 180, 240, 360 and 1440 minutes, as shown in \Cref{eq:delta_vol}
    \item Binary indicators for the day of the week (Monday, Tuesday, \dots , Saturday). 
    \item Binary indicators for the time of day (Hour\_00, Hour\_01, \dots, Hour\_22).  
\end{itemize}

\subsection{Baselines}
To compare and assess the potential improvement of the proposed pipeline and the utilization of the forecasts of other instruments, we consider a number of baselines, which are conventional machine learning methods. We include the Ordinary Least Squares (OLS), the LASSO model, and the Principal Component Regression (PCR). These simple methods are necessary to ensure that more complex methods do not overfit the data, especially in the case of forecasting returns. For the LASSO model, we choose the regularization hyperparameter using 10-fold cross-validation, while for PCR, we select the minimum number of principal components that retains 90\% of the variance.

Among ensemble models, we consider Random Forests as well as XGBOOST, as both methods are widely used in a variety of applications. For XGBOOST, we use the default values of the 'xgboost' package in Python, while for Random Forests, we determine the number of trees using cross-validation.

We also consider an Artificial Neural Network (ANN) and an LSTM model as simpler deep learning alternatives. Both of these models follow the same architecture as the GCN model we use. The LSTM model is identical to the GCN-LSTM architecture but without the GCN layer, while the ANN is the same as the LSTM model, without the LSTM layer.

%Both of these models are which are incorporated into the pipeline we use are simpler deep learning methods, which lack the propagation step. 

Lastly, we consider a naive forecast. For returns, this is equivalent to holding a long position in the product considered during all trading periods. For volatility, we set the naive forecast value equal to the observed value of the previous period. For differences in trading volume, we set all naive forecasts equal to zero.

\subsection{Implementation Details}
The general architecture of GCN-LSTM is shown in \Cref{fig:full_diagram}. The model is implemented using the tensorflow library in Python. A summary of the model's components is provided in \Cref{tab:gcn_details}. 

\begin{table}[] 
\small
\centering
\begin{tabular}{llll}
Layer Name                                                          & Layer Units                                                   & Activation Function                                    & Other parameters                                        \\ \hline
\textbf{LSTM}                                                       & 64                                                            & tanh                                                   & Sequence Length = 12                                    \\
\textbf{Dense Layer 1}                                              & 32                                                            & tanh                                                   &                                                         \\
\textbf{Dense Layer 2}                                              & 16                                                            & tanh                                                   &                                                         \\
\textbf{GCN}                                                        & 12 Graphs                                     & tanh                                                      &                                                         \\
\textbf{Output Layer}                                               & 1                                                             & lienar                                                 &                                                         \\ \hline
\multicolumn{4}{l}{\textbf{Regularization}: L1 with parameter $10^{-5}$}                                                                                                                                                      \\
\multicolumn{4}{l}{\begin{tabular}[c]{@{}l@{}}\textbf{Training}: ADAM with step size equal to 5e-04 (other parameters set to default)\\ \quad \quad \quad \quad 120 epochs are used for the initial training, and 25 epochs used each time the window is rolled.\end{tabular}} \\
\textbf{Lookback Window}: 20000                                    &                                                               &                                                        &                                                         \\
\multicolumn{4}{l}{\textbf{Rolling Window Length}: 1500}                                                                                                                                                                                           \\ \hline
\end{tabular} \caption{Details for the GCN-LSTM used for the forecasting of Returns, Volatility and Trading Volume.} \label{tab:gcn_details}
\end{table}

\section{Results} \label{sec:results}
\subsection{Forecasting Returns} \label{sec:ret_results}
The results for forecasting the returns of ES-Mini and VX futures are shown in \Cref{tab:returns_spx_res,tab:returns_vix_res}, respectively. It should be noted that for returns, we consider data points for training only where there is sufficient liquidity (more than 1) and a ``reasonable" spread\footnote{We consider a spread ``reasonable" if it is below 15 basis points for the SPX futures, and 25 basis points for the VIX futures, respectively.} for at least one of ES\_1 or VX\_1. In addition, for the evaluation of each individual product, we only consider data points with sufficient liquidity and a ``reasonable" spread for each.

\begin{figure}[h]
   \includegraphics[trim=3.7cm 1cm 3.8cm 1.8cm, clip, ,width=1\textwidth]{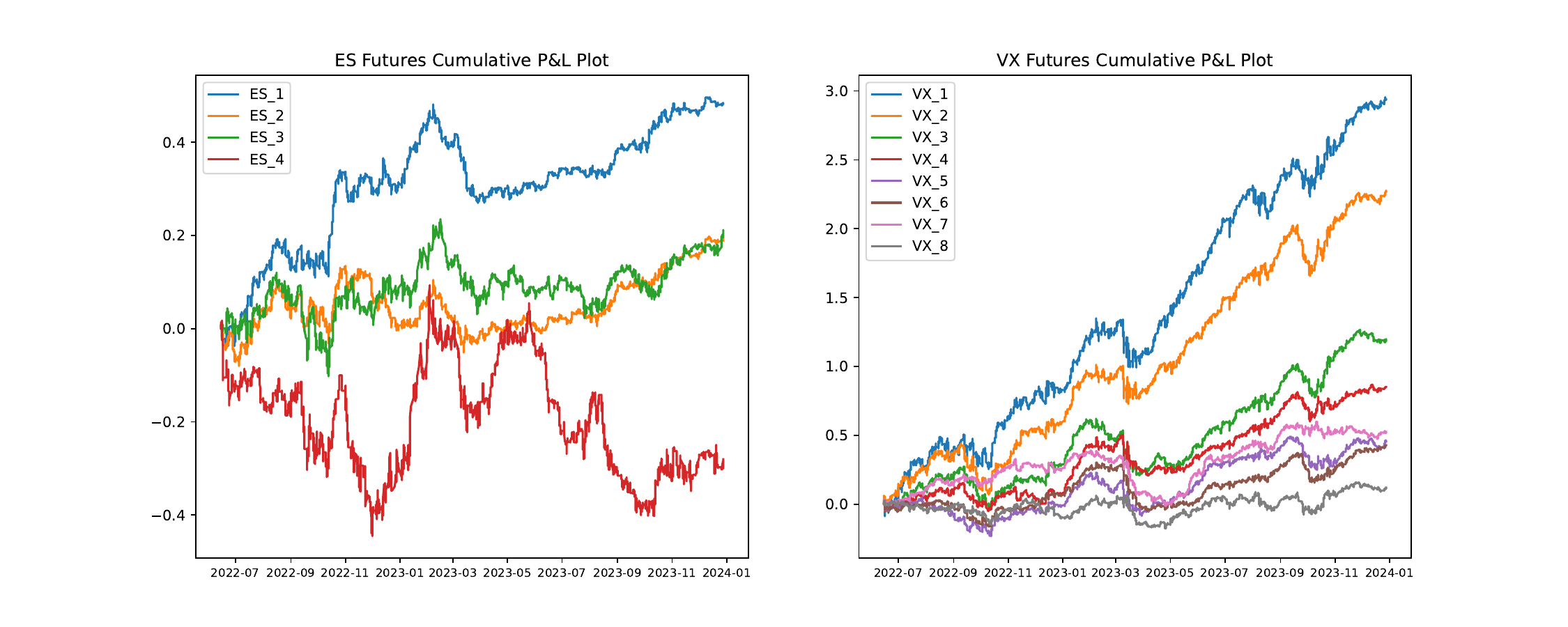} 
   \caption{Resulting cumulative P\&L plot for the hourly return forecasting problem.} 
   \label{fig:cumsum}
\end{figure}

It can be clearly seen that deep learning methods outperform the non-deep learning methods for ES\_1 and ES\_2 but are outperformed by linear methods for ES\_3 and ES\_4. As the number of points used for the evaluation of ES\_3 and ES\_4 is small, these results are less robust. GCN-LSTM is the best-performing deep learning method for ES\_1-ES\_3, and the results are very similar across the products, demonstrating the effectiveness of propagating forecasts among the different products. Importantly, GCN-LSTM is the highest-performing method among all methods for ES\_1 and ES\_2. The cumulative P\&L plot is shown in \Cref{fig:cumsum}.

For the VX futures, the GCN-LSTM method achieves the highest Sharpe Ratio for six out of eight products, scoring the second highest for the other two. Its PPD is either the highest or second highest for four out of eight scores, while in all cases it is above nine. Furthermore, for all products, GCN-LSTM outperforms ANN and LSTM in both SR and PPD, clearly demonstrating the improvement in forecasts resulting from including the GCN layer in our architecture. It should be noted that in most cases, LASSO and PCR both outperform LSTM and ANN, and they also outperform GCN-LSTM in terms of PPD.

In \Cref{tab:different_horizons}, we change the setting by considering different horizons. Even though for these cases, trainable points are fewer, results can still be promising. For the most liquid products (ES\_1, VX\_1, and VX\_2), there are a number of cases where we attain SR values higher than 1.5. The highest performances are generally observed for the daily results, with 9 out of 12 products resulting in SR values higher than 1. 

\begin{table}[] 
\small
\centering
\begin{tabular}{c|cc|cc|cc|cc|}  
Returns            & \multicolumn{2}{c|}{ES\_1}                                   & \multicolumn{2}{c|}{ES\_2}                                   & \multicolumn{2}{c|}{ES\_3}                                   & \multicolumn{2}{c|}{ES\_4}                                   \\
`Tradable' Periods & \multicolumn{2}{c|}{11236}                                   & \multicolumn{2}{c|}{2487}                                    & \multicolumn{2}{c|}{287}                                     & \multicolumn{2}{c|}{37}                                      \\
                   & SR                           & PPD                           & SR                           & PPD                           & SR                           & PPD                           & SR                           & PPD                           \\ \hline
Naive              & 0.051                        & 0.188                         & 0.054                        & 0.198                         & 0.047                        & 0.219                         & -0.012                       & -0.079                        \\
LM                 & 0.109                        & 0.401                         & -0.512                       & -1.912                        & -0.1                         & -0.583                        & 0.08                         & 0.73                          \\
LASSO              & 0.372                        & 1.336                         & 0.432                        & 1.605                         & {\color[HTML]{FE0000} 2.686} & {\color[HTML]{3166FF} 16.342} & {\color[HTML]{FE0000} 4.288} & {\color[HTML]{FE0000} 42.011} \\
LPCA               & 0.438                        & 1.582                         & 0.449                        & 1.727                         & {\color[HTML]{3166FF} 2.675} & {\color[HTML]{FE0000} 16.349} & {\color[HTML]{3166FF} 4.214} & {\color[HTML]{3166FF} 40.744} \\
Random Forests     & 0.663                        & 2.391                         & -0.235                       & -0.855                        & -0.062                       & -0.329                        & 0.398                        & 3.265                         \\
XGBoost            & 0.191                        & 0.718                         & -0.33                        & -1.227                        & -0.019                       & -0.104                        & 0.128                        & 1.018                         \\
ANN                & 1.048                        & 7.823                         & 0.913                        & 7.102                         & 0.918                        & 6.827                         & 0.545                        & 4.196                         \\
LSTM               & {\color[HTML]{3166FF} 1.281} & {\color[HTML]{3166FF} 9.591}  & {\color[HTML]{3166FF} 1.082} & {\color[HTML]{3166FF} 8.109}  & 0.947                        & 8.285                         & 1.301                        & 9.735                         \\
GCN-LSTM           & {\color[HTML]{FE0000} 1.465} & {\color[HTML]{FE0000} 10.941} & {\color[HTML]{FE0000} 1.383} & {\color[HTML]{FE0000} 9.865} & 1.341                        & 10.773                        & 1.283                        & 9.471                        
\end{tabular}
\caption{Predictive performance of different methods for hourly return forecasting of SPX futures. Here, SR and PPD correspond to Sharpe Ratio and Profit per Dollar respectively. In red (blue) we highlight the highest (second highest) performing method per product.} \label{tab:returns_spx_res}
\end{table}

\begin{table}[]
\resizebox{\textwidth}{!}{
\begin{tabular}{l|ll|ll|ll|ll|ll|ll|ll|ll}
Returns          & \multicolumn{2}{l|}{VX\_1}                                   & \multicolumn{2}{l|}{VX\_2}                                   & \multicolumn{2}{l|}{VX\_3}                                   & \multicolumn{2}{l|}{VX\_4}                                   & \multicolumn{2}{l|}{VX\_5}                                   & \multicolumn{2}{l|}{VX\_6}                                            & \multicolumn{2}{l|}{VX\_7}                                   & \multicolumn{2}{l|}{VX\_8}                                                        \\
Tradable Periods & \multicolumn{2}{l|}{6182}                                    & \multicolumn{2}{l|}{6243}                                    & \multicolumn{2}{l|}{5222}                                    & \multicolumn{2}{l|}{4459}                                    & \multicolumn{2}{l|}{3798}                                    & \multicolumn{2}{l|}{3277}                                             & \multicolumn{2}{l|}{792}                                     & \multicolumn{2}{l}{54}                                                            \\
                 & SR                           & PPD                           & SR                           & PPD                           & SR                           & PPD                           & SR                           & PPD                           & SR                           & PPD                           & SR                                    & PPD                           & SR                           & PPD                           & SR                           & \multicolumn{1}{l|}{PPD}                           \\ \hline
Naive            & -0.782                       & -12.094                       & -0.455                       & -5.273                        & -0.461                       & -4.048                        & -0.281                       & -2.073                        & -0.324                       & -2.13                         & -0.477                                & -2.776                        & -0.443                       & -2.666                        & -0.418                       & \multicolumn{1}{l|}{-2.899}                        \\
LM               & 0.339                        & 5.217                         & 0.542                        & 6.344                         & 0.162                        & 1.455                         & 0.045                        & 0.353                         & 0.329                        & 2.136                         & 0.323                                 & 1.962                         & 0.358                        & 2.219                         & -0.237                       & \multicolumn{1}{l|}{-1.686}                        \\
LASSO            & 0.71                         & {\color[HTML]{3166FF} 10.901} & 0.651                        & 7.653                         & {\color[HTML]{FE0000} 1.527} & {\color[HTML]{FE0000} 14.282} & 0.754                        & 5.896                         & 1.075                        & {\color[HTML]{FE0000} 14.243} & 1.099                                 & 9.882                         & 1.52                         & {\color[HTML]{3166FF} 15.72}  & {\color[HTML]{3166FF} 1.629} & \multicolumn{1}{l|}{11.847}                        \\
LPCA             & {\color[HTML]{3166FF} 1.202} & {\color[HTML]{FE0000} 18.43}  & 1.16                         & {\color[HTML]{FE0000} 14.157} & 1.362                        & {\color[HTML]{3166FF} 12.305} & 1.269                        & {\color[HTML]{3166FF} 9.916}  & 1.196                        & {\color[HTML]{3166FF} 13.237} & 1.195                                 & {\color[HTML]{FE0000} 12.437} & {\color[HTML]{FE0000} 1.562} & {\color[HTML]{FE0000} 16.044} & 1.459                        & \multicolumn{1}{l|}{{\color[HTML]{3166FF} 12.118}} \\
Random Forests   & 0.69                         & 10.854                        & 0.343                        & 4.024                         & 0.19                         & 1.692                         & 0.095                        & 0.742                         & 0.223                        & 1.519                         & -0.101                                & -0.628                        & -0.0                         & -0.001                        & 0.422                        & \multicolumn{1}{l|}{3.167}                         \\
XGBOOST          & -0.074                       & -1.16                         & 0.512                        & 5.96                          & -0.003                       & -0.027                        & -0.061                       & -0.472                        & 0.647                        & 4.49                          & 0.234                                 & 1.389                         & -0.188                       & -1.232                        & -0.146                       & \multicolumn{1}{l|}{-1.067}                        \\
ANN              & 0.729                        & 5.564                         & 0.588                        & 4.278                         & 0.59                         & 4.405                         & 0.933                        & 7.166                         & 1.118                        & 8.396                         & 0.458                                 & 3.491                         & 0.171                        & 1.295                         & -0.127                       & \multicolumn{1}{l|}{-0.972}                        \\
LSTM             & 0.995                        & 7.415                         & {\color[HTML]{3166FF} 1.287} & 9.585                         & 1.24                         & 9.207                         & {\color[HTML]{3166FF} 1.302} & 9.658                         & {\color[HTML]{3166FF} 1.125} & 8.518                         & {\color[HTML]{3166FF} 1.326}          & 9.957                         & 1.291                        & 9.655                         & 1.197                        & \multicolumn{1}{l|}{9.042}                         \\
GCN-LSTM         & {\color[HTML]{FE0000} 2.488} & 62.657                        & {\color[HTML]{FE0000} 2.403} & {\color[HTML]{3166FF} 51.649} & {\color[HTML]{3166FF} 1.59} & 31.427                        & {\color[HTML]{FE0000} 1.496} & {\color[HTML]{FE0000} 21.859} & {\color[HTML]{FE0000} 1.095} & 14.430                        & {\color[HTML]{FE0000} \textit{1.037}} & {\color[HTML]{3166FF} 11.495} & {\color[HTML]{3166FF} 1.259} & 9.050                         & {\color[HTML]{FE0000} 0.339} & \multicolumn{1}{l|}{{\color[HTML]{FE0000} 3.816}}
\end{tabular}}
\vspace{1mm}
\caption{Predictive performance of different methods for hourly return forecasting of VX futures. Here, SR and PPD correspond to Sharpe Ratio and Profit per Dollar respectively. In red (resp., blue) we highlight the highest (resp., second highest) performing method for each product.} \label{tab:returns_vix_res}
\end{table}

\subsection{Forecasting of Volatility}
\label{sec:volatility_results}
The results for forecasting the realized volatility of ES-Mini and VX futures are shown in \Cref{tab:vol_spx} and \Cref{tab:vol_vix}, respectively. In this case, deep learning methods outperform the non-deep learning methods for all products considered in terms of QLIKE. In most cases, GCN-LSTM and LSTM have similar performance, with the former generally outperforming the latter. GCN-LSTM is the best-performing method for all SPX products and also for 5 out of 7 products for VIX, being the second-best performer in the other two cases.

Results are quite different when considering the HMSE metric. In this case, deep learning methods appear to perform worse than many of the other considered methods. The best-performing method appears to be Random Forests, being the only one to outperform the naive forecasts.\footnote{While not reported here for brevity, results using the Mean Squared Error as an evaluation metric appear to give similar results to HMSE.}

There appears to be a trade-off between achieving high QLIKE and high HMSE. QLIKE shows the ability of a model to correctly identify the subsequent levels of volatility, while HMSE focuses on the ability of models to have relatively low deviations from the actual predictability. This distinction makes it unclear whether the improvement is significant.

Thus, as GCN-LSTM seems to outperform LSTM for 9 out of 11 products in terms of QLIKE and for 8 out of 11 products in terms of HMSE, the propagation step seems to result in an improvement.

\begin{table}[]
\small
\centering
\begin{tabular}{l|ll|ll|ll|ll}
      & \multicolumn{2}{l|}{3-Hour} & \multicolumn{2}{l|}{4-Hour} & \multicolumn{2}{l|}{6-Hour} & \multicolumn{2}{l}{Day} \\
      & SR           & PPD          & SR           & PPD          & SR           & PPD          & SR         & PPD        \\ \hline
ES\_1 & 0.717        & 5.196        & 1.649        & 13.064       & 1.074        & 7.233        & 1.599      & 9.269      \\
ES\_2 & -0.088       & -0.654       & 1.301        & 9.366        & 0.665        & 4.777        & 1.98       & 11.509     \\
ES\_3 & -0.436       & -4.186       & 1.45         & 12.395       & -0.066       & -0.551       & 1.591      & 12.288     \\
ES\_4 & -0.426       & -5.692       & -0.587       & -7.789       & -0.883       & -9.906       & -0.404     & -3.694     \\
VX\_1 & -1.158       & -28.283      & 2.936        & 74.197       & 1.219        & 28.802       & 2.245      & 40.572     \\
VX\_2 & 3.225        & 63.552       & 2.033        & 39.333       & 1.949        & 35.612       & 1.232      & 18.686     \\
VX\_3 & -0.370       & -5.898       & 2.562        & 41.811       & 2.313        & 35.868       & 0.74       & 9.623      \\
VX\_4 & -0.595       & -8.055       & 0.454        & 6.619        & 1.459        & 20.54        & 1.032      & 12.019     \\
VX\_5 & -2.547       & -31.665      & 2.115        & 28.18        & 2.146        & 26.684       & 1.299      & 13.704     \\
VX\_6 & -1.184       & -12.915      & 3.093        & 35.465       & 0.708        & 7.438        & 2.387      & 23.377     \\
VX\_7 & -0.657       & -7.909       & 0.182        & 2.131        & 0.589        & 6.651        & 0.948      & 9.677      \\
VX\_8 & -1.995       & -23.842      & -0.538       & -7.012       & 1.748        & 21.197       & 1.286      & 14.79      \\ \hline
\end{tabular}  \caption{This table shows the predictive performance for return forecasting at different horizons.}  \label{tab:different_horizons}
\end{table}

\begin{table}[] 
\small
\centering
\begin{tabular}{c|cc|cc|cc|cc|} 
Volatility     & \multicolumn{2}{c|}{ES\_1}                                        & \multicolumn{2}{c|}{ES\_2}                                        & \multicolumn{2}{c|}{ES\_3}                                        & \multicolumn{2}{c|}{ES\_4}                                        \\
               & QLIKE                           & HMSE                            & QLIKE                           & HMSE                            & QLIKE                           & HMSE                            & QLIKE                           & HMSE                            \\ \hline
Naive          & 7.68e-01                        & {\color[HTML]{FE0000} 3.60e-11} & 1.27e+00                        & {\color[HTML]{3166FF} 5.80e-11} & 3.20e+01                        & {\color[HTML]{3166FF} 2.35e-08} & 3.74e+03                        & {\color[HTML]{3166FF} 7.54e-08} \\
LM             & 7.39e-01                        & 2.38e-09                        & 1.38e+00                        & 2.07e-09                        & 1.43e+01                        & 9.54e-08                        & 2.13e+06                        & 1.38e-03                        \\
LASSO          & 7.35e-01                        & 9.77e-10                        & 1.38e+00                        & 8.10e-11                        & 1.16e+01                        & 3.10e-07                        & 2.78e+02                        & 1.28e-05                        \\
LPCA           & 1.24e+00                        & 9.12e-09                        & 1.38e+00                        & 1.33e-09                        & 9.49e+00                        & {\color[HTML]{FE0000} 1.88e-08} & 3.37e+02                        & 8.03e-03                        \\
Random Forests & 2.56e+00                        & {\color[HTML]{3166FF} 4.20e-11} & 1.71e+00                        & {\color[HTML]{FE0000} 5.20e-11} & 1.45e+01                        & 1.60e-08                        & 1.98e+02                        & {\color[HTML]{FE0000} 5.02e-08} \\
XGBOOST        & 1.79e+00                        & 1.93e-10                        & 1.54e+00                        & 7.86e-10                        & 8.04e+00                        & 9.33e-06                        & 2.42e+04                        & 5.24e-03                        \\
ANN            & 3.95e-07                        & 3.20e-06                        & 2.51e-05                        & 1.86e-04                        & 7.01e-07                        & 5.60e-06                        & 3.37e-07                        & 2.74e-06                        \\
LSTM           & {\color[HTML]{3166FF} 1.30e-08} & 1.02e-07                        & {\color[HTML]{3166FF} 3.02e-09} & 2.00e-08                        & {\color[HTML]{3166FF} 1.47e-07} & 1.13e-06                        & {\color[HTML]{3166FF} 1.82e-07} & 1.45e-06                        \\
GCN-LSTM       & {\color[HTML]{FE0000} 1.00e-09} & 1.20e-08                        & {\color[HTML]{FE0000} 3.00e-09} & 2.10e-08                        & {\color[HTML]{FE0000} 1.27e-07} & 9.75e-07                        & {\color[HTML]{FE0000} 5.00e-08} & 3.98e-07                       
\end{tabular} \caption{Predictive performance of different methods for hourly volatility forecasting for SPX futures. Here, QLIKE and HMSE correspond to Quasi-likelihood and Heteroskedasticity Adjusted Mean Squared Error. In red (blue) we highlight the highest (second highest) performing method per product.} \label{tab:vol_spx}
\end{table}

\begin{table}[] 
\resizebox{\textwidth}{!}{
\begin{tabular}{l|ll|ll|ll|ll|ll|ll|ll|}
Volatility     & \multicolumn{2}{l|}{VX\_1}                                        & \multicolumn{2}{l|}{VX\_2}                                        & \multicolumn{2}{l|}{VX\_3}                                        & \multicolumn{2}{l|}{VX\_4}                                        & \multicolumn{2}{l|}{VX\_5}                                        & \multicolumn{2}{l|}{VX\_6}                                        & \multicolumn{2}{l|}{VX\_7}                                        \\
               & \multicolumn{1}{c}{QLIKE}       & \multicolumn{1}{c|}{HMSE}       & \multicolumn{1}{c}{QLIKE}       & \multicolumn{1}{c|}{HMSE}       & \multicolumn{1}{c}{QLIKE}       & \multicolumn{1}{c|}{HMSE}       & \multicolumn{1}{c}{QLIKE}       & \multicolumn{1}{c|}{HMSE}       & \multicolumn{1}{c}{QLIKE}       & \multicolumn{1}{c|}{HMSE}       & \multicolumn{1}{c}{MSE}         & \multicolumn{1}{c|}{MAE}        & \multicolumn{1}{c}{QLIKE}       & \multicolumn{1}{c|}{HMSE}       \\ \hline
Naive          & 1.74e+00                        & {\color[HTML]{3166FF} 9.45e-08} & 2.75e+00                        & {\color[HTML]{3166FF} 5.96e-08} & 3.24e+00                        & {\color[HTML]{3166FF} 3.68e-08} & 3.76e+00                        & {\color[HTML]{3166FF} 2.82e-08} & 3.79e+00                        & {\color[HTML]{3166FF} 2.20e-08} & 3.28e+00                        & {\color[HTML]{3166FF} 1.50e-08} & 5.34e+00                        & 2.58e-08                        \\
LM             & 1.93e+00                        & 4.76e-07                        & 2.81e+00                        & 5.91e-05                        & 3.09e+00                        & 1.76e-07                        & 3.31e+00                        & 7.09e-08                        & 3.61e+00                        & 6.48e-08                        & 3.05e+00                        & 3.50e-08                        & 5.17e+00                        & 6.57e-06                        \\
LASSO          & 1.94e+00                        & 5.82e-07                        & 2.84e+00                        & 6.84e-05                        & 3.17e+00                        & 2.35e-07                        & 3.31e+00                        & 5.98e-08                        & 3.68e+00                        & 1.28e-07                        & 3.10e+00                        & 5.04e-08                        & 5.02e+00                        & 1.28e-05                        \\
LPCA           & 1.59e+00                        & 7.94e-06                        & 2.20e+00                        & 3.76e-05                        & 2.26e+00                        & 7.97e-03                        & 2.36e+00                        & 7.49e-05                        & 2.23e+00                        & 9.59e-07                        & 2.07e+00                        & 3.15e-07                        & 2.51e+00                        & {\color[HTML]{3166FF} 2.16e-08} \\
Random Forests & 9.84e-01                        & {\color[HTML]{FE0000} 6.10e-08} & 1.53e+00                        & {\color[HTML]{FE0000} 3.69e-08} & 1.83e+00                        & {\color[HTML]{FE0000} 2.30e-08} & 1.90e+00                        & {\color[HTML]{FE0000} 1.89e-08} & 2.03e+00                        & {\color[HTML]{FE0000} 1.63e-08} & 1.82e+00                        & {\color[HTML]{FE0000} 1.01e-08} & 2.76e+00                        & {\color[HTML]{FE0000} 1.51e-08} \\
XGBOOST        & 1.45e+00                        & 1.88e-07                        & 2.79e+00                        & 1.79e-06                        & 1.70e+00                        & 9.37e-07                        & 2.13e+00                        & 2.17e-06                        & 2.13e+00                        & 9.90e-07                        & 2.01e+00                        & 6.82e-07                        & 2.51e+00                        & 8.07e-05                        \\
ANN            & 6.75e-07                        & 5.34e-06                        & 2.08e-07                        & 1.67e-06                        & 2.56e-07                        & 2.04e-06                        & 1.78e-07                        & 1.43e-06                        & 3.38e-07                        & 2.86e-06                        & 2.96e-07                        & 2.38e-06                        & 7.10e-07                        & 5.71e-06                        \\
LSTM           & {\color[HTML]{3166FF} 1.27e-07} & 1.01e-06                        & {\color[HTML]{3166FF} 1.01e-07} & 8.04e-07                        & {\color[HTML]{3166FF} 4.00e-08} & 3.16e-07                        & {\color[HTML]{FE0000} 3.10e-08} & 2.52e-07                        & {\color[HTML]{3166FF} 1.80e-08} & 1.48e-07                        & {\color[HTML]{3166FF} 1.20e-08} & 9.20e-08                        & {\color[HTML]{FE0000} 2.10e-08} & 1.71e-07                        \\
GCN-LSTM       & {\color[HTML]{FE0000} 1.15e-07} & 9.14e-07                        & {\color[HTML]{FE0000} 5.50e-08} & 4.37e-07                        & {\color[HTML]{FE0000} 2.40e-08} & 1.92e-07                        & {\color[HTML]{3166FF} 8.30e-08} & 6.65e-07                        & {\color[HTML]{FE0000} 1.30e-08} & 1.02e-07                        & {\color[HTML]{FE0000} 9.00e-09} & 7.20e-08                        & {\color[HTML]{3166FF} 1.65e-07} & 1.31e-06                       
\end{tabular}} \caption{Predictive performance of different methods for  hourly volatility forecasting for VIX futures. Here, QLIKE and HMSE correspond to Quasi-likelihood and Heteroskedasticity Adjusted Mean Squared Error. In red (blue) we highlight the highest (second highest) performing method per product.} \label{tab:vol_vix}
\end{table}

\subsection{Forecasting Trading Volume}
\label{sec:volume_results}
The results for forecasting ES-Mini and VX trading volume are shown in \Cref{tab:volume_spx} and \Cref{tab:volume_vix}, respectively. In terms of MAE, GCN-LSTM is the best-performing method for 3 out of 4 SPX products and is either the best or second-best method for 5 out of 8 VIX products. Importantly, it is the best-performing method for the three most liquid products: ES\_1, VX\_1, and VX\_2. It outperforms LSTM for 3 out of 4 SPX products and the 3 VIX products closer to expiration, but is outperformed by LSTM for 5 out of 8 VIX products.

For MSE, the results differ significantly, with GCN-LSTM not being the best-performing method for any product. This outcome is potentially due to the use of MAE as the loss function for this architecture, which seems to be outperformed by other methods that tend to produce forecasts with lower variance (closer to the center). In fact, both LSTM and GCN-LSTM are outperformed by the naive models in a number of cases.

\begin{table}[H]
\small
\centering
\begin{tabular}{l|ll|ll|ll|ll|}
Volume         & \multicolumn{2}{l|}{ES\_1}                                          & \multicolumn{2}{l|}{ES\_2}                                          & \multicolumn{2}{l|}{ES\_3}                                          & \multicolumn{2}{l|}{ES\_4}                                          \\
               & MAE                              & MSE                              & MAE                              & MSE                              & MAE                              & MSE                              & MAE                              & MSE                              \\ \hline
Naive          & 1.617e-03                        & 6.625e-01                        & 5.171e-03                        & 8.256e-01                        & {\color[HTML]{3166FF} 4.954e-03} & 8.556e-01                        & 9.175e-04                        & 2.989e-01                        \\
LM             & 1.370e-02                        & 4.189e-01                        & 6.184e-02                        & {\color[HTML]{FE0000} 6.146e-01} & 1.056e-01                        & {\color[HTML]{343434} 6.926e-01} & 3.967e-02                        & 2.390e-01                        \\
LASSO          & 8.385e+00                        & 8.473e+00                        & 5.379e+00                        & 5.978e+00                        & 9.656e-01                        & 1.518e+00                        & 4.688e-02                        & 3.238e-01                        \\
LPCA           & 2.000e-03                        & 4.190e-01                        & 1.260e-01                        & 6.640e-01                        & 1.140e-01                        & 8.020e-01                        & 6.000e-03                        & 2.390e-01                        \\
Random Forests & 3.278e-02                        & {\color[HTML]{3166FF} 3.679e-01} & 8.158e-04                        & {\color[HTML]{3166FF} 6.332e-01} & 6.896e-02                        & 7.376e-01                        & 1.204e-02                        & 2.242e-01                        \\
XGBOOST        & 2.285e-02                        & {\color[HTML]{FE0000} 3.902e-02} & 4.226e-03                        & 6.954e-01                        & 3.734e-02                        & 7.611e-01                        & 1.826e-02                        & 2.297e-01                        \\
ANN            & {\color[HTML]{3166FF} 3.951e-04} & 7.509e-01                        & {\color[HTML]{FE0000} 5.713e-05} & 9.547e-01                        & 6.464e-03                        & {\color[HTML]{FE0000} 6.758e-01} & 1.705e-03                        & {\color[HTML]{FE0000} 2.037e-01} \\
LSTM           & 1.200e-03                        & 7.520e-01                        & {\color[HTML]{3166FF} 2.786e-04} & 9.568e-01                        & 6.766e-03                        & {\color[HTML]{3166FF} 6.795e-01} & {\color[HTML]{3166FF} 6.861e-04} & {\color[HTML]{3166FF} 2.090e-01} \\
GCN-LSTM       & {\color[HTML]{FE0000} 1.733e-04} & 7.518e-01                        & 2.992e-04                        & 9.589e-01                        & {\color[HTML]{FE0000} 4.118e-04} & 6.859e-01                        & {\color[HTML]{FE0000} 9.797e-05} & {\color[HTML]{3166FF} 2.090e-01}
\end{tabular} \caption{Predictive performance of different methods for hourly trading volume forecasting for SPX futures. Here, MAE and MAE correspond to the Mean Absolute Error and Mean Squared Error respectively. In red (blue) we highlight the highest (second highest) performing method per product.} \label{tab:volume_spx}
\end{table}

\begin{table}[H]
\resizebox{\textwidth}{!}{
\begin{tabular}{l|ll|ll|ll|ll|ll|ll|ll|ll|}
Volume         & \multicolumn{2}{l|}{VX\_1}                                          & \multicolumn{2}{l|}{VX\_2}                                          & \multicolumn{2}{l|}{VX\_3}                                          & \multicolumn{2}{l|}{VX\_4}                                         & \multicolumn{2}{l|}{VX\_5}                                         & \multicolumn{2}{l|}{VX\_6}                                         & \multicolumn{2}{l|}{VX\_7}                                         & \multicolumn{2}{l|}{VX\_8}                                          \\
               & MAE                              & MSE                              & MAE                              & MSE                              & MAE                              & MSE                              & MAE                              & MSE                             & MAE                              & MSE                             & MAE                              & MSE                             & MAE                              & MSE                             & MAE                              & MSE                              \\ \hline
Naive          & 6.532e-04                        & 1.175e+00                        & {\color[HTML]{3531FF} 1.175e-03} & 1.301e+00                        & {\color[HTML]{FE0000} 1.788e-04} & 1.386e00                         & 5.923e-04                        & 1.490e00                        & 6.767e-03                        & 1.659e00                        & 6.188e-03                        & 1.640e+00                       & 5.030e-03                        & 1.581e00                        & 1.432e-03                        & 1.271e00                         \\
LM             & 3.148e-02                        & 9.600e-01                        & 3.676e-02                        & 9.616e-01                        & 2.776e-02                        & {\color[HTML]{3166FF} 1.005e00}  & 4.056e-02                        & {\color[HTML]{FE0000} 1.097e00} & 4.741e-02                        & {\color[HTML]{3166FF} 1.188e00} & 4.545e-02                        & {\color[HTML]{FE0000} 1.162e00} & 3.882e-02                        & {\color[HTML]{FE0000} 1.021e00} & 6.887e-02                        & 8.977e-01                        \\
LASSO          & 5.786e+00                        & 6.138e+00                        & 5.449e+00                        & 5.894e+00                        & 4.314e00                         & 4.874e00                         & 3.625e00                         & 4.278e00                        & 2.905e00                         & 3.733e00                        & 2.440e00                         & 3.361e00                        & 1.918e+00                        & 2.882e00                        & 1.490e00                         & 2.368e00                         \\
LPCA           & 2.100e-02                        & 9.000e-01                        & 4.400e-02                        & 9.650e-01                        & 1.100e-02                        & 1.083e00                         & 4.300e-02                        & 1.246e00                        & 2.000e-03                        & 1.366e00                        & 3.600e-02                        & 1.338e00                        & 2.100e-02                        & 1.202e00                        & 0.000e00                         & 1.027e00                         \\
Random Forests & 5.763e-02                        & {\color[HTML]{FE0000} 7.550e-01} & 3.673e-02                        & {\color[HTML]{FE0000} 8.570e-01} & 9.401e-03                        & {\color[HTML]{FE0000} 9.759e-01} & 1.610e-02                        & {\color[HTML]{3166FF} 1.124e00} & 3.573e-03                        & {\color[HTML]{FE0000} 1.183e00} & 5.550e-03                        & {\color[HTML]{3166FF} 1.227e00} & 6.956e-03                        & 1.144e00                        & 9.397e-02                        & {\color[HTML]{3166FF} 8.967e-01} \\
XGBOOST        & 3.257e-02                        & {\color[HTML]{3166FF} 7.920e-01} & 1.413e-02                        & {\color[HTML]{3166FF} 9.162e-01} & 2.750e-02                        & 1.021e00                         & 5.050e-02                        & 1.147e00                        & 8.728e-02                        & 1.309e00                        & 3.588e-03                        & 1.243e00                        & 3.490e-02                        & {\color[HTML]{3166FF} 1.281e00} & 1.467e-01                        & 9.500e-01                        \\
ANN            & {\color[HTML]{343434} 6.158e-04} & 1.253e00                         & 2.799e-03                        & 1.369e00                         & 3.338e-03                        & 1.412e00                         & 1.475e-03                        & 1.479e00                        & {\color[HTML]{FE0000} 5.859e-04} & 1.526e00                        & {\color[HTML]{FE0000} 1.203e-03} & 1.503e00                        & {\color[HTML]{FE0000} 2.044e-03} & 1.415e00                        & 2.099e-03                        & {\color[HTML]{FE0000} 1.126e-03} \\
LSTM           & {\color[HTML]{3531FF} 2.583e-04} & 1.256e+00                        & {\color[HTML]{343434} 2.106e-03} & 1.370e+00                        & 1.684e-03                        & 1.417e00                         & {\color[HTML]{FE0000} 1.232e-04} & 1.482e00                        & {\color[HTML]{3166FF} 9.453e-04} & 1.529e00                        & {\color[HTML]{3166FF} 1.960e-03} & 1.506e00                        & {\color[HTML]{3166FF} 2.265e-03} & 1.424e00                        & {\color[HTML]{FE0000} 3.685e-05} & 1.139e00                         \\
GCN-LSTM       & {\color[HTML]{FE0000} 1.378e-04} & 1.256e00                         & {\color[HTML]{FE0000} 9.283e-04} & 1.370e00                         & {\color[HTML]{3531FF} 1.120e-03} & 1.417e00                         & {\color[HTML]{3166FF} 4.977e-04} & 1.482e00                        & 1.225e-03                        & 1.529e00                        & 2.081e-03                        & 1.507e00                        & 3.075e-03                        & 1.417e00                        & {\color[HTML]{3166FF} 2.692e-04} & 1.139e00                        
\end{tabular}} \caption{Predictive performance of different methods for \textbf{hourly trading volume} forecasting for VIX futures. MAE and MAE correspond to the Mean Absolute Error and Mean Squared Error respectively. In red (blue) we highlight the highest (second highest) performing method per product.} \label{tab:volume_vix}
\end{table}

\subsection{Feature Importance}
While it can be challenging to extract feature importance scores for deep learning models, the LASSO coefficients obtained during the baseline step act as a readily available and interpretable feature importance metric. Across the 14 instruments,\footnote{For trading volume, there are only 12 products, as the indices are not tradable.} we obtained the median rank of each feature in terms of the absolute value of the fitted coefficients.

For the return forecasting task, only 6 of the features are deemed significant for more than half the cases, as return forecasting is inherently a high error-to-signal ratio problem. These features include the 30, 240, and 360-minute returns ($R_t^{30-\textrm{minute}},R_t^{240-\textrm{minute}},R_t^{360-\textrm{minute}}$), the 30-minute negative semi-variance ($\textrm{RV-neg}_t^{30-\textrm{minute}}$), as well as the exponentially weighted return and realized volatility with a parameter equal to 0.75 ($R_t^{(0.75)-\textrm{weighted}}$,$\textrm{RV}_t^{(0.75)-\textrm{weighted}}$).

The exponentially weighted realized volatilities (weigh\_rv) appear to be among the 20 most important features for all three tasks. For volatility, all weigh\_rv features are included in the list of most important features, with weigh\_rv\_1.0 having the highest median rank and being a non-trivial predictor for 11 products. The 10 and 60-minute realized volatilities are also in the list of the most important features.

For trading volume, some of the most important features unsurprisingly include the preceding trading volumes, with the 90-minute feature scoring the highest median rank. The time-of-day feature also scores high for trading volume and volatility, with many binary hour indicator features after 12:00 appearing in the top 20. Lastly, most days are included in the top 20 for trading volume, with the Monday binary indicator scoring highly for volatility.

\begin{table}[]
\resizebox{\textwidth}{!}{
\begin{tabular}{lllllll}
                   & \multicolumn{1}{c}{Returns}  &                            & \multicolumn{1}{c}{Volatility}        &                                    & \multicolumn{1}{c}{Volume} &                            \\
                   & Median Rank                  & Non-zero Coefficients      & Median Rank                           & Non-zero Coefficients              & Median Rank                          & Non-zero Coefficients      \\
OFI\_360           & 1.0                          & 4                          & \cellcolor[HTML]{FE996B}47.0          & \cellcolor[HTML]{FE996B}9          & 43.0                                 & 12                         \\
OFI\_1440          & 1.0                          & 3                          & 1.0                                   & 6                                  & 19.0                                 & 12                         \\
weigh\_rets\_0.75  & \cellcolor[HTML]{FE996B}90.0 & \cellcolor[HTML]{FE996B}9  & 30.5                                  & 7                                  & 35.5                                 & 12                         \\
weigh\_rets\_0.9   & 1.0                          & 2                          & 30.5                                  & 7                                  & 44.0                                 & 11                         \\
weigh\_rets\_0.975 & 1.0                          & 1                          & 1.0                                   & 2                                  & 45.0                                 & 11                         \\
weigh\_rets\_0.99  & 1.0                          &                            & 1.0                                   & 6                                  & 22.5                                 & 8                          \\
weigh\_rets\_0.999 & 1.0                          & 2                          & 21.5                                  & 7                                  & 46.5                                 & 12                         \\
weigh\_rets\_1.0   & 1.0                          & 1                          & 15.0                                  & 7                                  & 41.5                                 & 10                         \\
weigh\_rv\_0.75    & \cellcolor[HTML]{FE996B}15.5 & \cellcolor[HTML]{FE996B}7  & \cellcolor[HTML]{FE996B}87.0          & \cellcolor[HTML]{FE996B}13         & 35.0                                 & 12                         \\
weigh\_rv\_0.9     & 1.0                          & 5                          & \cellcolor[HTML]{FE996B}78.0          & \cellcolor[HTML]{FE996B}8          & 63.5                                 & 12                         \\
weigh\_rv\_0.975   & 1.0                          & 4                          & \cellcolor[HTML]{FE996B}75.5          & \cellcolor[HTML]{FE996B}8          & \cellcolor[HTML]{FE996B}83.0         & \cellcolor[HTML]{FE996B}12 \\
weigh\_rv\_0.99    & 1.0                          & 1                          & \cellcolor[HTML]{FE996B}\textbf{75.5} & \cellcolor[HTML]{FE996B}10         & \cellcolor[HTML]{FE996B}76.0         & \cellcolor[HTML]{FE996B}11 \\
weigh\_rv\_0.999   & 1.0                          & 3                          & \cellcolor[HTML]{FE996B}82.5          & \cellcolor[HTML]{FE996B}8          & 64.0                                 & 11                         \\
weigh\_rv\_1.0     & 1.0                          & 5                          & \cellcolor[HTML]{FE996B}91.0          & \cellcolor[HTML]{FE996B}11         & 54.5                                 & 11                         \\
ret\_5        & 1.0                          & 5                          & 28.0                                  & 7                                  & 17.5                                 & 12                         \\
ret\_10       & 1.0                          & 3                          & 15.5                                  & 7                                  & 20.0                                 & 11                         \\
ret\_30       & \cellcolor[HTML]{FE996B}32.0 & \cellcolor[HTML]{FE996B}7  & \cellcolor[HTML]{FE996B}61.5          & \cellcolor[HTML]{FE996B}10         & 17.5                                 & 11                         \\
ret\_60       & 1.0                          & 6                          & 1.0                                   & 6                                  & 33.5                                 & 11                         \\
ret\_90       & 1.0                          & 5                          & 41.0                                  & 10                                 & 21.5                                 & 12                         \\
ret\_180      & 1.0                          & 3                          & 1.0                                   & 5                                  & 20.0                                 & 10                         \\
ret\_240      & \cellcolor[HTML]{FE996B}83.5 & \cellcolor[HTML]{FE996B}9  & \cellcolor[HTML]{FE996B}45.0          & \cellcolor[HTML]{FE996B}8          & 16.0                                 & 11                         \\
ret\_360      & \cellcolor[HTML]{FE996B}89.5 & \cellcolor[HTML]{FE996B}12 & 36.0                                  & 8                                  & 23.0                                 & 11                         \\
ret\_1440     & 1.0                          & 1                          & 1.0                                   & 6                                  & 15.0                                 & 11                         \\
rv\_5         & 1.0                          & 1                          & 36.0                                  & 7                                  & 1.0                                  & 6                          \\
rv\_10        & 1.0                          & 2                          & \cellcolor[HTML]{FE996B}\textit{76.0} & \cellcolor[HTML]{FE996B}\textit{9} & 3.5                                  & 7                          \\
rv\_30        & 1.0                          & 2                          & 1.0                                   & 4                                  & 10.0                                 & 7                          \\
rv\_60        & 1.0                          & 0                          & \cellcolor[HTML]{FE996B}83.5          & \cellcolor[HTML]{FE996B}9          & 1.0                                  & 6                          \\
rv\_90        & 1.0                          & 3                          & 1.0                                   & 3                                  & 9.0                                  & 8                          \\
rv\_pos360    & 1.0                          & 3                          & 1.0                                   & 3                                  & 33.5                                 & 10                         \\
rv\_pos1440   & 1.0                          & 3                          & \cellcolor[HTML]{FE996B}75.5          & \cellcolor[HTML]{FE996B}10         & 19.0                                 & 8                          \\
rv\_neg5      & 1.0                          & 1                          & 22.5                                  & 7                                  & 17.5                                 & 11                         \\
rv\_neg10     & 1.0                          & 3                          & 18.0                                  & 7                                  & 14.0                                 & 7                          \\
rv\_neg30     & \cellcolor[HTML]{FE996B}87.0 & \cellcolor[HTML]{FE996B}8  & 1.0                                   & 5                                  & 23.5                                 & 11                 F        \\
num\_trades\_5     & 1.0                          & 3                          & 1.0                                   & 3                                  & 49.5                                 & 12                         \\
num\_trades\_10    & 1.0                          & 1                          & 1.0                                   & 2                                  & \cellcolor[HTML]{FE996B}74.0         & \cellcolor[HTML]{FE996B}12 \\
num\_trades\_30    & 1.0                          & 1                          & 1.0                                   & 5                                  & \cellcolor[HTML]{FE996B}95.0         & \cellcolor[HTML]{FE996B}12 \\
num\_trades\_60    & 1.0                          & 2                          & 28.0                                  & 7                                  & \cellcolor[HTML]{FE996B}84.0         & \cellcolor[HTML]{FE996B}12 \\
num\_trades\_90    & 1.0                          & 1                          & 1.0                                   & 3                                  & \cellcolor[HTML]{FE996B}91.0         & \cellcolor[HTML]{FE996B}11 \\
num\_trades\_180   & 1.0                          & 1                          & 1.0                                   & 3                                  & \cellcolor[HTML]{FE996B}73.5         & \cellcolor[HTML]{FE996B}12 \\
num\_trades\_240   & 1.0                          & 0                          & 1.0                                   & 2                                  & 43.0                                 & 12                         \\
num\_trades\_360   & 1.0                          & 1                          & 1.0                                   & 2                                  & \cellcolor[HTML]{FE996B}84.5         & \cellcolor[HTML]{FE996B}12 \\
num\_trades\_1440  & 1.0                          & 2                          & 1.0                                   & 5                                  & \cellcolor[HTML]{FE996B}85.0         & \cellcolor[HTML]{FE996B}12 \\
Monday             & 1.0                          & 5                          & \cellcolor[HTML]{FE996B}64.5          & \cellcolor[HTML]{FE996B}10         & 64.5                                 & 12                         \\
Tuesday            & 1.0                          & 1                          & 1.0                                   & 6                                  & \cellcolor[HTML]{FE996B}81.5         & \cellcolor[HTML]{FE996B}12 \\
Wednesday          & 1.0                          & 1                          & 19.5                                  & 7                                  & \cellcolor[HTML]{FE996B}78.0         & \cellcolor[HTML]{FE996B}12 \\
Thursday           & 1.0                          & 1                          & 31.5                                  & 8                                  & \cellcolor[HTML]{FE996B}78.5         & \cellcolor[HTML]{FE996B}12 \\
Friday             & 1.0                          & 3                          & 42.5                                  & 8                                  & \cellcolor[HTML]{FE996B}76.0         & \cellcolor[HTML]{FE996B}12 \\
Saturday           & 1.0                          & 0                          & 1.0                                   & 0                                  & 1.0                                  & 0                          \\
Hour\_11           & 1.0                          & 3                          & 1.0                                   & 6                                  & 56.0                                 & 12                         \\
Hour\_12           & 1.0                          & 4                          & \cellcolor[HTML]{FE996B}60.5          & \cellcolor[HTML]{FE996B}9          & \cellcolor[HTML]{FE996B}69.5         & \cellcolor[HTML]{FE996B}12 \\
Hour\_13           & 1.0                          & 4                          & \cellcolor[HTML]{FE996B}82.0          & \cellcolor[HTML]{FE996B}12         & \cellcolor[HTML]{FE996B}84.5         & \cellcolor[HTML]{FE996B}12 \\
Hour\_14           & 1.0                          & 2                          & \cellcolor[HTML]{FE996B}82.5          & \cellcolor[HTML]{FE996B}11         & \cellcolor[HTML]{FE996B}75.5         & \cellcolor[HTML]{FE996B}12 \\
Hour\_15           & 1.0                          & 6                          & \cellcolor[HTML]{FE996B}63.0          & \cellcolor[HTML]{FE996B}12         & 64.0                                 & 12                         \\
Hour\_16           & 1.0                          & 3                          & 41.0                                  & 8                                  & 62.5                                 & 12                         \\
Hour\_17           & 1.0                          & 1                          & 44.0                                  & 8                                  & 62.0                                 & 12                         \\
Hour\_18           & 1.0                          & 3                          & \cellcolor[HTML]{FE996B}53.5          & \cellcolor[HTML]{FE996B}10         & \cellcolor[HTML]{FE996B}68.5         & \cellcolor[HTML]{FE996B}12 \\
Hour\_19           & 1.0                          & 6                          & \cellcolor[HTML]{FE996B}72.0          & \cellcolor[HTML]{FE996B}10         & \cellcolor[HTML]{FE996B}76.0         & \cellcolor[HTML]{FE996B}12 \\
Hour\_20           & 1.0                          & 6                          & 42.5                                  & 8                                  & 63.0                                 & 12                         \\
Hour\_21           & 1.0                          & 5                          & \cellcolor[HTML]{FE996B}69.5          & \cellcolor[HTML]{FE996B}12         & \cellcolor[HTML]{FE996B}88.5         & \cellcolor[HTML]{FE996B}12 \\
Hour\_22           & 1.0                          & 3                          & 16.5                                  & 7                                  & \cellcolor[HTML]{FE996B}71.5         & \cellcolor[HTML]{FE996B}12
\end{tabular}}
\caption{Feature important metric extracted from LASSO models for the three major forecasting problems studied. In red, we show the 20 features with the highest median rank. For presentation purposes, we exclude some features which are not highly important for any of the three problems (the excluded Features include: The OFI features below 240, realized volatility of 180 minutes or more, positive realized volatility of between 5 and 240 minutes, negative realized volatility of above 90 minutes, and hours 1-10.)}
\end{table}

\subsection{Different Configurations}
In \Cref{tab:robustness}, we show the results obtained by GCN-LSTM for ES\_1 and VX\_1 for each forecasted quantity. These results are compared with alternative configurations of the same model. In particular, instead of using 12 graphs, we use the Contemporaneous Weighted, Contemporaneous Unweighted, Lagged Weighted, and Lagged Unweighted separately (3 graphs in each case). We further explore different loss functions, including the MAE, MSE, and SR loss functions (where applicable). We also consider a configuration with `Non-parallel modules', meaning that the weights for the LSTM and dense layers are shared across nodes. Note that in the previous subsections, we also considered the LSTM and ANN models, which were simplifications of GCN-LSTM without the GCN layer. We draw the overall conclusions for the different tasks considered.

\bb \; For returns, we clearly observe that using different loss functions yields significantly worse results. Using all the graphs appears to be beneficial as it outperforms all other cases where we only use a subset, with Contemporaneous Unweighted being the best and second-best performing for ES\_1 and VX\_1, respectively, among the four types of graphs.  

\bb \; For volatility, the proposed setup achieves the highest performance in terms of QLIKE for VX\_1 and the second-best performance for ES\_1 (outperformed by Contemporaneous Weighted). Using MSE as the loss function clearly diminishes performance, while using MAE appears to perform almost comparably to the QLIKE loss. 

\bb \; For volume, the proposed setup seems to be outperformed by the configuration where weights are shared across nodes, which is most likely a natural outcome of higher correlation among nodes. Furthermore, the lagged networks appear to mostly outperform the contemporaneous networks.

\begin{table}[]
\resizebox{\textwidth}{!}{
\centering
\begin{tabular}{l|llll|llll|llll}
                           & \multicolumn{4}{l|}{Returns}                                                                                                & \multicolumn{4}{l|}{Volatility}                                                                                                           & \multicolumn{4}{l}{Volume}                                                                                                                \\ \cline{2-13} 
                           & \multicolumn{2}{l}{ES\_1}                                    & \multicolumn{2}{l|}{VIX\_1}                                  & \multicolumn{2}{l}{ES\_1}                                           & \multicolumn{2}{l|}{VIX\_1}                                         & \multicolumn{2}{l}{ES\_1}                                           & \multicolumn{2}{l}{VIX\_1}                                          \\
                           & SR                           & PPD                           & SR                           & PPD                           & QLIKE                            & HMSE                             & QLIKE                            & HMSE                             & MSE                              & MAE                              & MSE                             & MAE                               \\ \hline
Contemporaneous Weighted   & 0.766                        & {\color[HTML]{6434FC} 6.910}  & {\color[HTML]{6434FC} 0.786} & {\color[HTML]{6434FC} 10.013} & {\color[HTML]{FE0000} 1.699e-10} & {\color[HTML]{FE0000} 3.397e-10} & 1.342e-07                        & {\color[HTML]{6434FC} 2.683e-07} & 7.524e-01                        & 1.745e-03                        & {\color[HTML]{FE0000} 1.256e00} & 7.950e-04                         \\
Contemporaneous Unweighted & {\color[HTML]{6434FC} 1.184} & 6.484                         & 0.765                        & 5.955                         & 7.694e-08                        & 1.539e-07                        & 1.659e-07                        & 3.317e-07                        & 7.524e-01                        & 2.366e-03                        & {\color[HTML]{FE0000} 1.256e00} & 2.784e-03                         \\
Lagged Weighted            & 0.510                        & 2.694                         & 0.688                        & 8.895                         & 1.576e-04                        & 2.467e-04                        & 2.791e-04                        & 4.113e-04                        & 7.520e-01                        & 8.715e-04                        & {\color[HTML]{FE0000} 1.256e00} & 5.011e-04                         \\
Lagged Unweighted          & 0.640                        & 3.125                         & 0.615                        & 7.647                         & 4.697e-09                        & 9.397e-09                        & 5.422e-07                        & 1.084e-06                        & {\color[HTML]{6434FC} 7.519e-01} & 8.872e-04                        & {\color[HTML]{FE0000} 1.256e00} & 5.011e-04                         \\
Loss Function: MSE         & 0.368                        & 1.223                         & 0.332                        & 4.328                         & 5.902e-04                        & 1.811e-03                        & 6.666e-07                        & 1.353e-06                        & 7.719e-01                        & 4.707e-04                        & 1.271e00                        & 1.730e-03                         \\
Loss Function: MAE         & -0.138                       & -0.463                        & -0.316                       & -4.027                        & 2.296e-09                        & {\color[HTML]{6434FC} 4.591e-09} & {\color[HTML]{6434FC} 1.278e-07} & {\color[HTML]{FE0000} 2.554e-07} & NA                               & NA                               & NA                              & NA                                \\
Loss Function: SR          & 0.007                        & 0.023                         & -0.254                       & -3.272                        & NA                               & NA                               & NA                               & NA                               & NA                               & NA                               & NA                              & NA                                \\
Non-parallel modules       & 0.213                        & 0.710                         & 0.486                        & 6.281                         & 1.523e-07                        & 3.031-07                         & 1.740e-07                        & 3.459e-07                        & 7.706e-01                        & {\color[HTML]{FE0000} 1.420e-04} & {\color[HTML]{6434FC} 1.270e00} & {\color[HTML]{FE0000} 2.001ee-05} \\
Used model                 & {\color[HTML]{FE0000} 1.465} & {\color[HTML]{FE0000} 13.941} & {\color[HTML]{FE0000} 1.328} & {\color[HTML]{FE0000} 10.243} & {\color[HTML]{6434FC} 1.00e-09}  & 1.20e-08                         & {\color[HTML]{FE0000} 1.15e-07}  & 9.140e-07                        & {\color[HTML]{FE0000} 7.518e-01} & {\color[HTML]{6434FC} 1.733e-04} & {\color[HTML]{FE0000} 1.256e00} & {\color[HTML]{6434FC} 1.346e-04}  \\ \hline
\end{tabular}} \caption{Predictive performance of different methods for hourly return, volatility and volume forecasting for ES\_1 and VX\_1 futures. In red (blue) we highlight the highest (second highest) performing method per product.}  \label{tab:robustness}
\end{table}

\section{Conclusion} \label{sec:conclusion}
We have analyzed the interaction between the different E-mini S\&P 500 futures and Cboe Volatility Index futures. The analysis of the graphs in our study provided several key insights into the relationships between the different futures products, offering a visual representation of the correlation structure at play in this market. The clear clustering of SPX and VIX nodes, the prevalence of negative correlations between returns and volatility, and the varied influence of liquidity levels on the connectivity of different graphs all provide valuable insights and highlight the complexity of financial markets. These findings solidify the utility of establishing a cross-asset flow of information for the different forecasting tasks.  

By employing a novel architecture of the GCN-LSTM class, we aimed to enhance the predictability for a number of forecasting tasks. We have demonstrated that the proposed GCN-LSTM architecture provides significant improvements in forecasting accuracy over the base LSTM model. The forecasting results for returns in \Cref{sec:ret_results} indicate that deep learning methods, particularly GCN-LSTM, generally outperform traditional methods for more liquid products like ES\_1 and ES\_2. Similarly, \Cref{sec:volatility_results} and \Cref{sec:volume_results} show that the already high performance of the LSTM model for the volatility and volume forecasting tasks, respectively, is further improved by our GCN-LSTM model. 

Importantly, in this paper we integrate economically significant loss functions into our model. By using the Sharpe Ratio for financial returns and a quasi-likelihood loss function for volatility, we ensure that the predictions of our model are not only reliable but also economically meaningful. This aligns our forecasting objectives with practical financial decision-making, rendering our approach highly relevant for practitioners in the futures markets. 

While our study has provided valuable insights, it also opens up several avenues for future research. One potential direction is the exploration of other types of derivative products, such as options, where larger networks can be constructed as the strike price also varies. The combination of both futures and options data would result in an interesting examination of the resulting graphs, providing insights into the interactions between options and futures contracts. Another interesting angle would be to explore this methodology in a different market, such as the commodities futures market.

\paragraph{Acknowledgments:} 
This work was supported by the UK Engineering and Physical Sciences Research Council (EPSRC) grants EP/R513295/1 and EP/V520202/1. 

\bibliography{main}

\appendix

\end{document}